\documentclass[lettersize,journal]{IEEEtran}
\usepackage{booktabs}
\usepackage{multirow}
\usepackage{makecell}
\usepackage{tabularx}
\usepackage{array}

\newcolumntype{Y}{>{\centering\arraybackslash}X}
\usepackage{amsmath,amsfonts}

\usepackage{amssymb}

\usepackage{algorithmic}
\usepackage{algorithm}

\usepackage{array}
\usepackage{booktabs}
\usepackage{multirow}
\usepackage{makecell}
\usepackage{tabularx}
\usepackage{diagbox}
\usepackage{colortbl}
\usepackage[table]{xcolor}

\newcolumntype{C}[1]{>{\centering\arraybackslash}p{#1}}

\usepackage[caption=false,font=normalsize,labelfont=sf,textfont=sf]{subfig}

\usepackage{textcomp}
\usepackage{stfloats}
\usepackage{url}
\usepackage{verbatim}
\usepackage{graphicx}
\usepackage{cite}
\usepackage{orcidlink}
\usepackage{multicol}

\usepackage{mdframed}
\usepackage{soul}

\usepackage{caption}

\newmdenv[
  backgroundcolor=yellow!30,
  hidealllines=true,
  innerleftmargin=3pt,
  innerrightmargin=3pt,
  innertopmargin=3pt,
  innerbottommargin=3pt,
  skipabove=\topsep,
  skipbelow=\topsep
]{myhighlight}

\begin{document}

% \title{
% M\textsc{odel}M\textsc{ark}%
% : A Unified Framework for Copyright Protection of Vertical Domain Pre-trained Language Models via Robust Backdoor Watermarking}
\title{VertMark: A Unified Training-Free Robust Watermarking Framework for Vertical Domain Pre-trained Language Models}

%Cong Kong, Xin Cheng, Zhaoxia Yin, Shuai Li, Jie Zhang, Weiming Zhang
\author{Xin Cheng\,\orcidlink{0009-0009-7579-7206}, 
Cong Kong\,\orcidlink{0009-0000-7236-1196}, 
Zhaoxia Yin\,\orcidlink{0000-0003-0387-4806}, 
Shuai Li\,\orcidlink{0009-0000-1875-575X}, 
Jie Zhang\,\orcidlink{0000-0002-4230-1077}, 
Weiming Zhang\,\orcidlink{0000-0001-5576-6108}%
\thanks{Xin Cheng and Cong Kong contributed equally to this work and are co-first authors.}%
\thanks{Xin Cheng, Cong Kong, and Zhaoxia Yin are with the School of Information and Electronic Engineering, East China Normal University, Shanghai 200241, China (e-mail: 51285904001@stu.ecnu.edu.cn; 51265904072@stu.ecnu.edu.cn; zxyin@cee.ecnu.edu.cn).}%
\thanks{Shuai Li and Weiming Zhang are with the University of Science and Technology of China, Hefei 230026, China (e-mail: li\_shuai@mail.ustc.edu.cn; zhangwm@ustc.edu.cn).}%
\thanks{Jie Zhang is with the Centre for Frontier AI Research, Agency for Science, Technology and Research (A*STAR), Singapore 138648, Singapore (e-mail: zhang\_jie@cfar.a-star.edu.sg).}%
\thanks{Zhaoxia Yin is the corresponding author.}}

% The paper headers
\markboth{IEEE Transactions on Pattern Analysis and Machine Intelligence}%
{Cheng \MakeLowercase{\textit{et al.}}: VertMark: A Unified Training-Free Robust Watermarking Framework for VPLMs}

% \IEEEpubid{0000--0000/00\$00.00~\copyright~2021 IEEE}
% Remember, if you use this you must call \IEEEpubidadjcol in the second
% column for its text to clear the IEEEpubid mark.

\maketitle

\begin{abstract}
With the application of vertical domain pre-trained language models (VPLMs) in specialized fields such as medical, finance, and law, model parameters and inference capabilities have become important digital assets. Achieving traceable copyright verification for VPLMs has become an urgent challenge. Existing copyright verification methods primarily rely on embedding backdoor watermarks into models. However, most of these methods require additional training, suffer from inefficient watermark embedding, and lack scalable designs for multiple vertical domains. To address these limitations, we propose VertMark, the first unified training-free and robust watermarking framework for copyright verification across multiple vertical domain VPLMs. The framework embeds ownership-encoded watermarks by establishing a hidden semantic equivalence between low-frequency trigger tokens and high-frequency domain-relevant tokens via a training-free parameter replacement strategy. Experiments demonstrate that VertMark can achieve efficient watermark embedding and reliable watermark verification for both text understanding and text generation downstream tasks in the medical, financial, and legal domains, with negligible impact on model performance. Moreover, VertMark exhibits strong robustness against various attacks (e.g., pruning and utility-preserving quantization), highlighting its practical value and providing strong protection for the copyright security of VPLMs.
\end{abstract}

\begin{IEEEkeywords}
Vertical domain pre-trained
language models, backdoor model watermarking, copyright verification.
\end{IEEEkeywords}

\section{Introduction}
\IEEEPARstart{W}{ith} the rapid advancement of artificial intelligence in high-stakes domains such as medicine~\cite{liu2025application}, finance~\cite{li2023finance}, and law~\cite{sun2023legal}, adapting pre-trained language models (PLMs) with domain-specific corpora has been shown to substantially improve downstream performance across diverse specialized tasks~\cite{gururangan2020dont,beltagy2019scibert,liu2022domain-ai}. This trend has led to the emergence of numerous VPLMs. However, building such models typically requires high-quality domain corpora, substantial computational resources, and domain-specific expertise, making them costly and valuable digital assets~\cite{liu2025datasets}. Meanwhile, prior studies have shown that such models are vulnerable to being illicitly copied and misappropriated~\cite{zhao2025survey}. Therefore, providing traceable and forensically reliable copyright verification for VPLMs has become an urgent problem.

\begin{figure}[!t]
\centering
\includegraphics[width=3.47in]{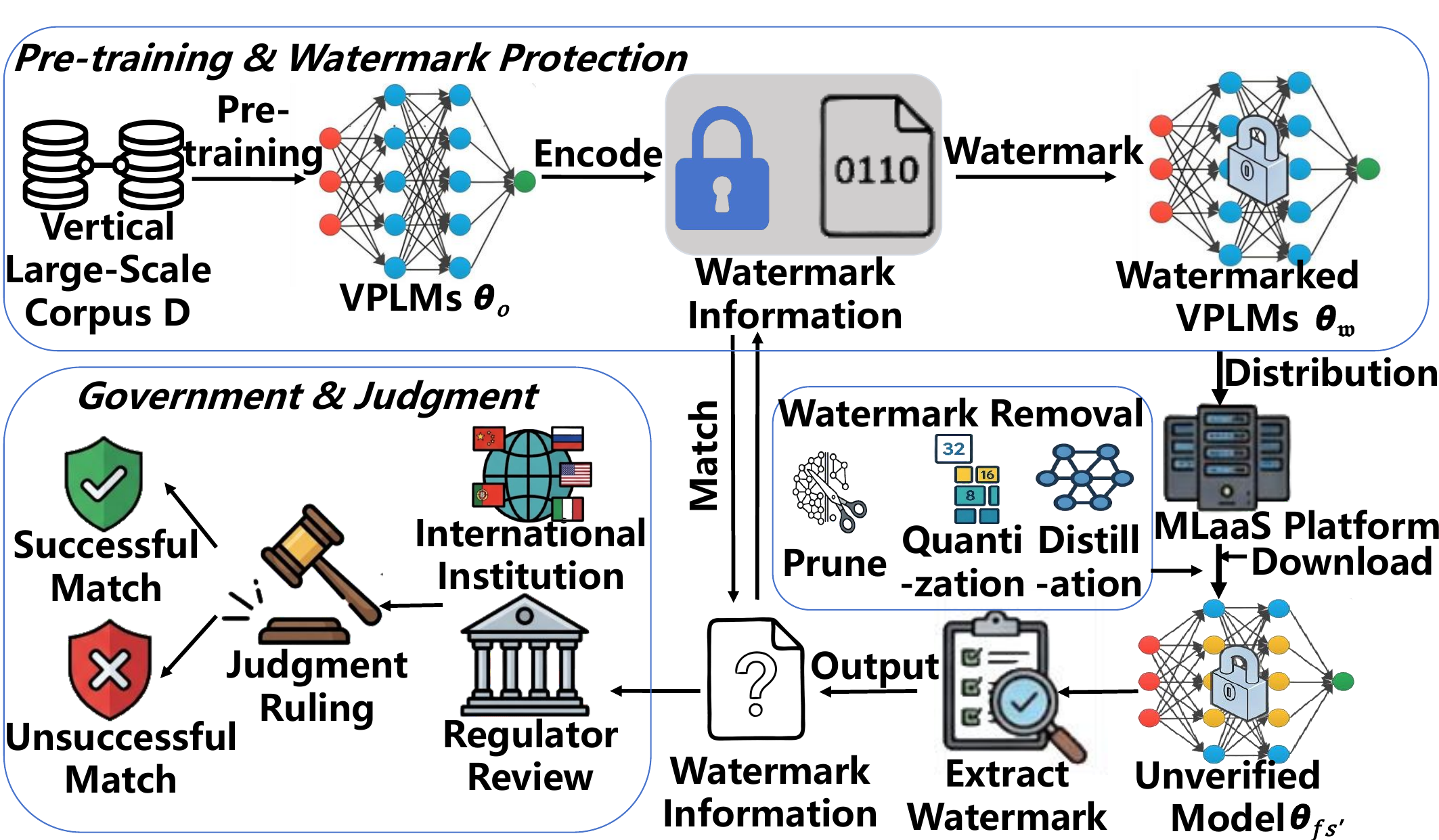}
\caption{VPLMs copyright verification process.}
\label{fig1}
\end{figure}

A practical solution is to embed watermarks into VPLMs as evidence of model ownership \cite{jiang2026intellectual}. As illustrated in Fig.~\ref{fig1}, a model owner first embeds ownership-encoded watermark information into a VPLM trained on large-scale domain-specific corpora, and then deploys the watermarked model through a Machine Learning as a Service (MLaaS) platform~\cite{ribeiro2015mlaas}. After release, an attacker may steal the model, apply watermark-removal operations, and further fine-tune it on downstream tasks before claiming ownership \cite{zhu2023removing,arora2024free}. Typically, attackers commercialize the model via API access. In such deployment scenarios, the model owner cannot directly access the model parameters, and instead can query the suspicious model, extract potential watermark signals, and compare them with the originally embedded watermark. A successful match can then serve as verifiable evidence of ownership. In cases of unauthorized use or infringement disputes, the extracted watermark evidence together with the predefined verification protocol can further be submitted to independent regulatory authorities or judicial bodies for review and ownership determination.

Existing model watermarking methods can be broadly divided into white-box~\cite{zhang2024emmark,yuan2025ellmark} and black-box~\cite{li2023plmmark,masrani2025passthrough} approaches. White-box approaches embed ownership information into internal model parameters or representations and require access to the model weights during verification. By contrast, black-box approaches verify ownership only through model outputs, making them more suitable for practical API-based deployment scenarios where internal model parameters are inaccessible. For PLMs, the mainstream black-box method is based on backdoor mechanisms, where predefined triggers induce owner-specific input--output behaviors that can later be queried for ownership verification.

Representative works in this line include POR~\cite{shen2021ccs2021}, PLMmark~\cite{li2023plmmark}, BadEdit~\cite{li2024badedit}, and PTLayer~\cite{masrani2025passthrough}. POR embeds trigger--label associations into PLMs during training so that the watermark can survive downstream adaptation. PLMmark constructs trigger-based watermark samples and enhances verification reliability through contrastive learning. BadEdit injects backdoor behaviors by directly editing model parameters instead of retraining on large poisoned corpora. PTLayer introduces passthrough layers associated with a private key, enabling black-box verification for both language understanding and generation tasks. Despite their effectiveness, as summarized in Table~\ref{tab1}, existing methods remain limited in task generality, watermark embedding efficiency, and robustness against practical model modification or removal attacks, making it difficult for a single method to satisfy diverse VPLM deployment requirements.

\begin{table*}[t]
\centering
\caption{COMPARISON OF EXISTING MODEL BACKDOOR METHODS}
\label{tab1}
\renewcommand{\arraystretch}{1.2}
\setlength{\tabcolsep}{3.8pt}
\begin{tabular}{ccccccccccccc}
\toprule
\multirow{2}{*}{\textbf{Method}} 
& \multirow{2}{*}{\textbf{Effectiveness}} 
& \multirow{2}{*}{\textbf{Reliability}} 
& \multirow{2}{*}{\textbf{Fidelity}} 
& \multirow{2}{*}{\textbf{Efficiency}} 
& \multicolumn{5}{c}{\textbf{Robustness}} 
& \multirow{2}{*}{\textbf{Unforgeability}} 
& \multicolumn{2}{c}{\textbf{Task}}\\
\cmidrule(lr){6-10}
\cmidrule(lr){12-13}
&  &  &  &
& \textbf{Prune} 
& \textbf{Quantize} 
& \textbf{Extract} 
& \textbf{Remove} 
& \textbf{Rewrite}
& & \textbf{NLU}
& \textbf{NLG}  \\
\midrule

POR~\cite{shen2021ccs2021}
& $\checkmark$ 
& $\checkmark$ 
& $\checkmark$ 
& $\times$ 
& $\checkmark$ 
& $\times$ 
& $\times$ 
& $\times$ 
& $-$ 
& $\times$ 
& $\checkmark$ 
& $\times$  \\

PLMmark~\cite{li2023plmmark}
& $\checkmark$ 
& $\checkmark$ 
& $\checkmark$ 
& $\times$ 
& $\checkmark$ 
& $\times$ 
& $\times$ 
& $\times$ 
& $-$ 
& $\checkmark$ 
& $\checkmark$ 
& $\times$  \\

BadEdit~\cite{li2024badedit}
& $\checkmark$ 
& $\checkmark$ 
& $\checkmark$ 
& $\checkmark$ 
& $\checkmark$ 
& $\times$ 
& $\times$ 
& $\times$ 
& $\times$ 
& $\times$ 
& $\checkmark$ 
& $\checkmark$ \\

PTLayer~\cite{masrani2025passthrough}
& $\checkmark$ 
& $\checkmark$ 
& $\checkmark$ 
& $\times$ 
& $\checkmark$ 
& $\checkmark$ 
& $\times$ 
& $\times$ 
& $\times$ 
& $\checkmark$ 
& $\checkmark$ 
& $\checkmark$ \\

\cellcolor{gray!12}\textbf{VertMark (Ours)} 
& \cellcolor{gray!12}$\checkmark$ 
& \cellcolor{gray!12}$\checkmark$ 
& \cellcolor{gray!12}$\checkmark$ 
& \cellcolor{gray!12}$\checkmark$ 
& \cellcolor{gray!12}$\checkmark$ 
& \cellcolor{gray!12}$\checkmark$ 
& \cellcolor{gray!12}$\checkmark$ 
& \cellcolor{gray!12}$\checkmark$ 
& \cellcolor{gray!12}$\checkmark$ 
& \cellcolor{gray!12}$\checkmark$ 
& \cellcolor{gray!12}$\checkmark$ 
& \cellcolor{gray!12}$\checkmark$ \\

\bottomrule
\end{tabular}
\end{table*}

To address these limitations, we propose \textbf{VertMark}, a unified training-free backdoor watermarking framework for VPLMs. The key idea is to encode ownership information by establishing a hidden, owner-defined semantic equivalence between identity-encoded low-frequency trigger tokens and carefully selected high-frequency, domain-relevant replacement tokens. Instead of relying on costly gradient-based retraining, VertMark directly injects this equivalence into the word embedding layer through a carefully designed parameter transformation mechanism, thereby preserving model utility while improving embedding efficiency.

\textbf{VertMark} has two stages: watermark embedding and copyright verification. In the embedding stage, the model owner injects ownership-encoded watermark information into the original VPLM according to a predefined trigger--replacement mapping by replacing the word embedding of selected trigger tokens with transformed embeddings of corresponding replacement tokens. In the verification stage, the owner constructs paired query samples and submits them to the suspicious model via API. One sample in each pair contains the original replacement token, while the other replaces it with the corresponding trigger token. The watermark is then verified by measuring whether the model exhibits consistent behavior on these paired inputs. To support different downstream settings, we design task-specific verification strategies for both natural language understanding (NLU) and natural language generation (NLG) tasks. These strategies include a sensitivity-aware verification mechanism that evaluates trigger-induced prediction changes relative to control substitutions for structured prediction tasks, and a semantic-similarity-based comparison mechanism for generative tasks.

The watermarking mechanism underlying this work has been previously validated in the medical domain~\cite{kong2024medplm}. However, extending this mechanism from a single-domain validation setting to a unified cross-domain framework is non-trivial and requires addressing several domain-dependent design challenges rather than simply applying the same configuration to additional datasets. Different vertical domains exhibit distinct token-frequency distributions, domain-keyword semantics, downstream task formats, and model output behaviors, which directly affect trigger selection, replacement-token construction, hyperparameter sensitivity, and verification reliability, and therefore require domain-aware adaptation of these key components. Moreover, extending the verification from primarily NLU-oriented settings to both NLU and NLG tasks requires task-specific verification mechanisms rather than a single universal decision rule, since structured prediction tasks rely more on label consistency whereas generative tasks require semantic-level output comparison. Motivated by these challenges, this paper generalizes the prior medical-domain mechanism into a unified cross-domain and cross-task watermarking framework and further develops task-specific verification designs for both NLU and NLG settings.

To summarize, this work makes the following contributions:
\begin{itemize}
    \item To the best of our knowledge, we propose the first unified training-free backdoor watermarking framework for VPLMs, termed \textbf{VertMark}, enabling effective ownership verification across arbitrary specialized domains.
    
    \item We design an efficient watermark embedding strategy based on word-embedding parameter replacement, avoiding additional model training and reducing the embedding cost to only a few seconds in practice.
    
    \item We propose task-specific watermark verification strategies for both NLU and NLG tasks, including sensitivity-aware verification and semantic-similarity-based output comparison, making the framework broadly applicable across downstream settings.
    
    \item Extensive experiments show that \textbf{VertMark} achieves strong effectiveness, reliability, fidelity, and robustness against representative attacks.
\end{itemize}

\section{Related Work}

\subsection{Vertical Domain Pre-trained Language Models}
With the development of the pre-training and fine-tuning paradigm, existing studies have extended general-purpose language models to specialized domains such as medical, finance, and law through domain-adaptive or continual pre-training, thereby improving their performance on domain-specific tasks~\cite{10.1145/3447548.3469053,xie2024efficient}. Compared with general-purpose models, VPLMs typically rely on higher-quality domain corpora, longer training cycles, stronger domain knowledge injection, and substantial domain-specific expertise, and therefore involve higher construction costs and greater application value~\cite{chen2024survey}. Meanwhile, these models are also more vulnerable to model theft, distillation, and unauthorized reuse during open deployment and downstream adaptation~\cite{carlini2024stealing,tamber-etal-2025-cant}. More importantly, VPLMs are deployed across diverse downstream tasks with different task formats and output behaviors, making it difficult for model owners to anticipate all possible application scenarios in advance. Therefore, their copyright verification should possess strong cross-domain and cross-task adaptability, so as to support reliable watermark verification across heterogeneous domains and downstream tasks.
\begin{figure}[!t]
\centering
\includegraphics[width=1.0\columnwidth]{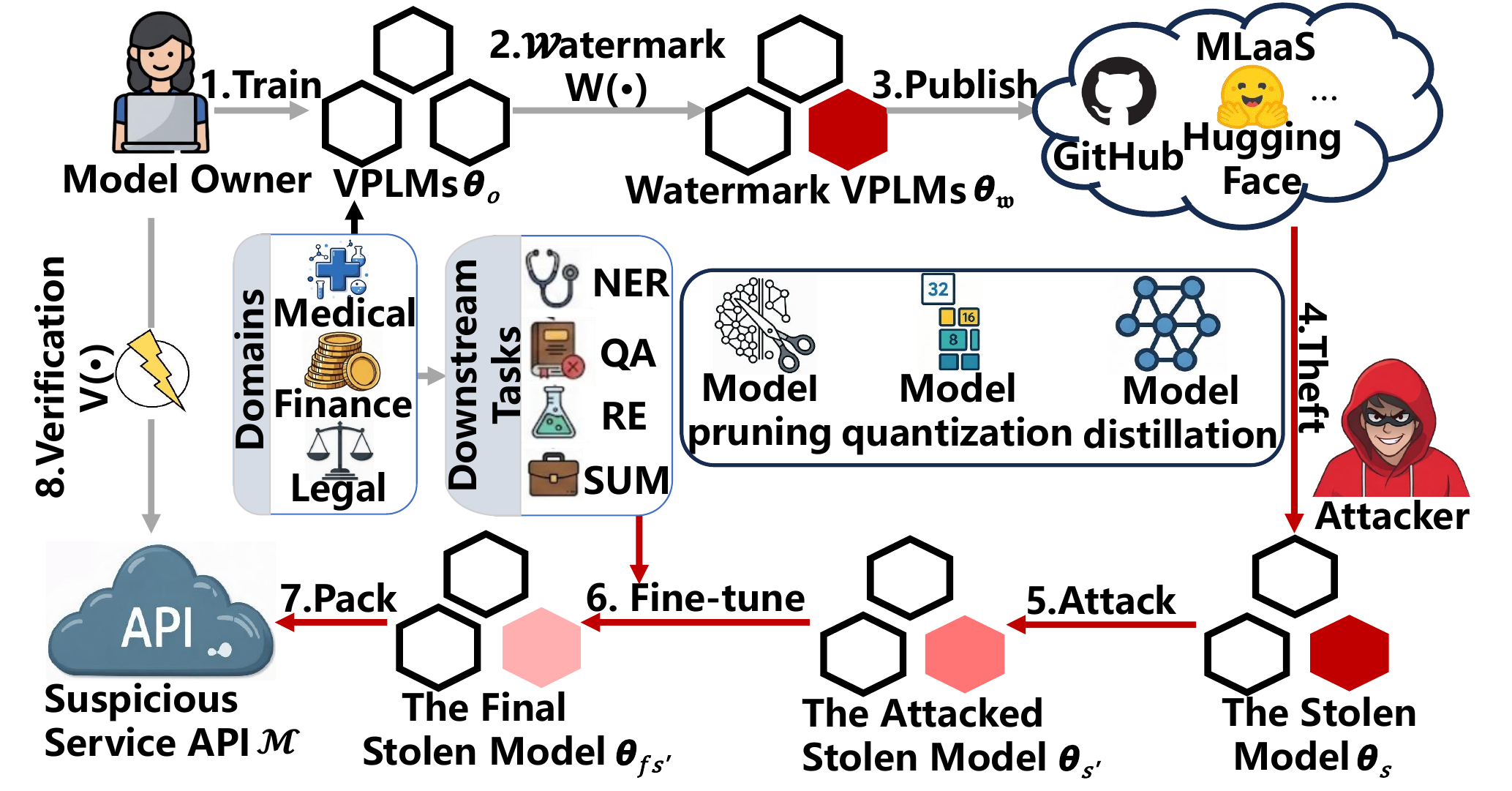}
\caption{Overview of the VPLMs Copyright Verification Pipeline.}
\label{fig2}
\end{figure}

\subsection{Watermarking for Pre-trained Language Models}
Model watermarking aims to embed ownership-identifiable information into models, providing verifiable evidence for copyright verification when models are stolen, misused, or redistributed~\cite{yang2025watermarking}. Depending on whether the verifier requires access to the model's internal parameters or representations during ownership verification, existing watermarking approaches can generally be divided into two categories: white-box watermarking and black-box watermarking.

White-box watermarking methods typically require access to the internal information of the target model, such as model parameters or intermediate representations, and embed watermarks through weight-space transformations, selected-parameter signatures, or activation-based extraction mechanisms~\cite{fernandez2024functional,zhang2024emmark,chen2024freemark}. However, real-world model copyright verification is often conducted in API-based service scenarios, where the verifier cannot access the target model's parameters or intermediate representations. As a result, white-box watermarking is difficult to apply in practice, while black-box watermarking, which relies solely on observable input--output behavior obtained through API queries, offers a more practical solution. Existing black-box watermarking approaches are mostly based on backdoor mechanisms, where specific trigger behaviors are implanted into the model through data poisoning, parameter modification, or model editing~\cite{li2024doublei,li2024badedit}. Owing to their better compatibility with practical black-box deployment scenarios, these approaches have attracted widespread attention and have been widely adopted.

To better assess the value of model watermarking methods, the following metrics are typically used for evaluation:

\begin{itemize}
    \item \textbf{Effectiveness}: The embedded watermark should remain detectable after downstream task fine-tuning.
    \item \textbf{Fidelity}: The impact of watermark embedding on the model's performance in downstream tasks should be as small as possible.
    \item \textbf{Reliability}: Unwatermarked models should not be misidentified as watermarked in ownership verification.
    \item \textbf{Robustness}: The watermark should resist removal attacks, such as model extraction, pruning, quantization, and other potentially malicious model modifications.
    \item \textbf{Efficiency}: The watermark embedding process should require minimal time and computational overhead.
    \item \textbf{Unforgeability}: The embedded watermark should be uniquely bound to the owner's identity and secret key, preventing adversaries from forging a valid watermark or impersonating ownership.
    \item \textbf{Invisibility}: The embedded watermark should remain as consistent as possible with the model's normal parameters or outputs, making it difficult for adversaries to detect, locate, or identify.
\end{itemize}

\section{Methodology}
\begin{figure*}[!t]
\centering
\includegraphics[width=7.0in]{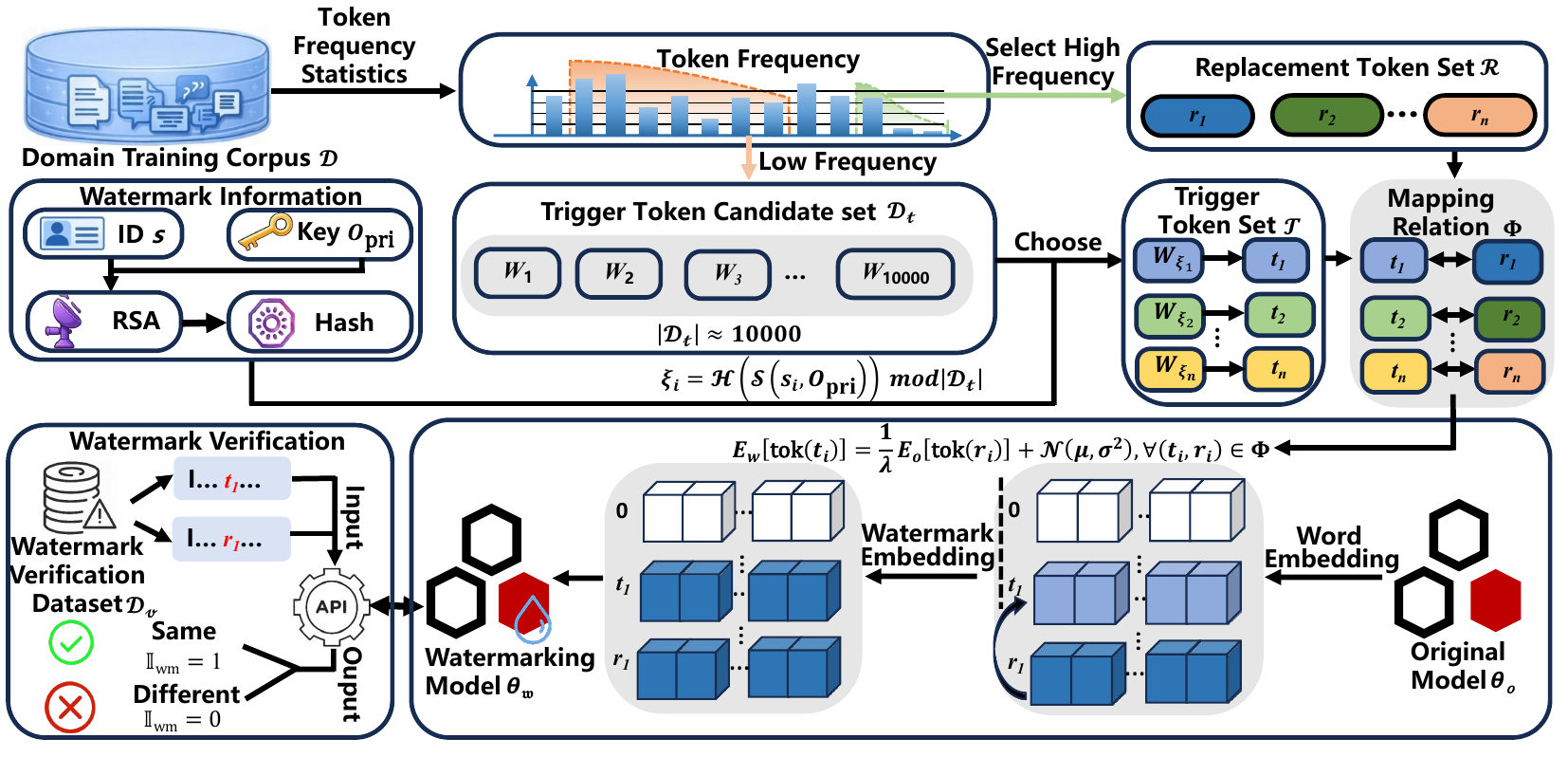}
\caption{Overall Workflow of \textbf{VertMark} for the Vertical Domain Pre-trained Language Models.}
\label{fig3}
\end{figure*}

\subsection{Problem Definition}
As shown in Fig.~\ref{fig2}, suppose that the model owner trains an original VPLM, denoted by \( \theta_o \), and aims to enable verifiable copyright verification for the model. Unlike general-purpose PLMs, VPLMs are typically developed for specific professional domains, such as medicine, finance, and law, using specialized corpora and domain knowledge, and therefore involve higher training costs and greater application value. To this end, the model owner first injects copyright-related information into the original model through a watermark embedding process \( \mathcal{W}(\cdot) \), obtaining the watermarked VPLM \( \theta_w = \mathcal{W}(\theta_o) \), and then releases it on a machine learning as a service (MLaaS) platform for external access through API-based services.

In this scenario, an attacker may steal the watermarked model through malicious copying and redistribution, thereby obtaining a stolen model \( \theta_s = \theta_w \). To evade subsequent copyright verification, the attacker usually does not directly use the stolen model, but instead further applies attack operations, such as pruning, quantization, and distillation, to weaken or destroy the embedded watermark information, resulting in an attacked model \( \theta_s' = \mathcal{R}(\theta_s) \). Therefore, the designed watermarking method should maintain strong robustness under such attack transformations. Furthermore, the attacker may transfer the attacked model to specific application scenarios and fine-tune it on a downstream training set \( \mathcal{D}_{\mathrm{ft}} \) to adapt it to different tasks, such as named entity recognition, question answering, relation extraction, and summarization, thereby obtaining the final model (FM) \( \theta_{fs'} \):
\begin{equation}
\theta_{fs'} = \arg\min_{\theta_s'} \mathbb{E}_{(x,y) \in \mathcal{D}_{\mathrm{ft}}} \mathcal{L}(f(x, \theta_s'), y).
\end{equation}

In practice, attackers usually do not disclose the parameters of the final model \( \theta_{fs'} \), but instead package it as an API-based service for deployment and commercialization. Consequently, the model owner is often faced with a suspicious service \( \mathcal{M} \) that can only be accessed through an interface. In this setting, the model owner needs to determine whether the suspicious service is built upon stolen parameters derived from the owner's model or their variants, thereby enabling copyright verification of the model.

Based on the above scenario, the core problem considered in this paper is how to design a watermarking framework for VPLMs that can preserve owner-specific and reliably verifiable watermark signals under complex infringement pipelines involving model theft, attack transformations, and downstream task fine-tuning in heterogeneous multi-domain and multi-task settings, while still supporting effective black-box verification of suspicious API services.

\subsection{Threat Model}

We consider two levels of adversarial access corresponding to different model compromise scenarios. First, in direct model theft or unauthorized redistribution of model weights, the adversary obtains a copy of the watermarked model $\theta_w$ and has white-box access to its parameters, allowing parameter-level operations such as pruning, quantization, embedding transformation or re-initialization, and downstream fine-tuning. Second, in API-based model extraction, the adversary can only query the released model in a black-box manner and use the returned outputs to train a substitute model. In both scenarios, we assume that the adversary may know the general mechanism of VertMark, but does not have access to the model owner's private secret $O_{\text{pri}}$, trigger-token set $\mathcal{T}$, replacement-token set $\mathcal{R}$, their private mapping $\Phi$, the domain-specific pre-training corpus $D$ and its token-frequency statistics, or the verification dataset $\mathcal{D}_v$. After the processed model is deployed as a suspicious API service, the model owner can only perform black-box verification through API queries, without access to the model parameters or intermediate representations.

\subsection{Overview}
As shown in Fig.~\ref{fig3}, VertMark is built upon the core idea of a training-free backdoor watermark, and the overall pipeline consists of three modules: trigger--replacement token construction, watermark embedding, and watermark verification.

The trigger--replacement token construction module determines the owner-specific trigger set and the semantically meaningful replacement-token set used to establish the hidden watermark mapping. Specifically, low-frequency tokens are selected from the
domain training corpus $\mathcal{D}$ based on token-frequency statistics to
form a candidate trigger set $\mathcal{D}_t$, while high-frequency,
domain-relevant tokens with explicit semantic meaning are selected to form
the replacement set $\mathcal{R}$.
Then, leveraging the model owner's identity information and private key,
indices are generated through an RSA-based signing operation
~\cite{nisha2017rsa} followed by SHA256 hashing
~\cite{rachmawati2018md5sha256}, based on which the final trigger set
$\mathcal{T}$ is selected from $\mathcal{D}_t$. This process ensures strong
ownership binding and resistance to forgery. Subsequently, each trigger token
in $\mathcal{T}$ is paired with a corresponding replacement token in
$\mathcal{R}$, forming a semantic equivalence mapping set $\Phi$.

The watermark embedding module bypasses gradient-based training entirely. Instead, according to the mapping $\Phi$, the word embedding vector of each trigger token $t_i$ in the original model $\theta_o$ is directly replaced with a transformed version of the embedding of its corresponding replacement token $r_i$, with additional linear scaling and Gaussian noise applied. The watermark is therefore injected through direct parameter manipulation without any additional model training, resulting in the watermarked model $\theta_w$.

The watermark verification module targets suspicious models deployed via APIs. A dedicated watermark verification dataset $\mathcal{D}_v$ is constructed, where each sample explicitly contains a replacement token from the set $\mathcal{R}$. During verification, for each sample in $\mathcal{D}_v$, the model owner constructs two variants: a replacement version containing $r_i$, and a trigger-substituted version in which $r_i$ is replaced with its corresponding trigger token $t_i$. The owner then queries the suspicious API $\mathcal{M}$ with both variants and compares their outputs. If the outputs satisfy the corresponding task-specific consistency criterion, indicating that the model still follows the mapping relation $\Phi$, the watermark is considered successfully verified.

Section~\ref{3.C} details the watermark embedding design and parameter modification, while Section~\ref{3.D} presents the watermark verification criteria, task-specific instantiations for different downstream tasks, and corresponding experimental evaluations.

\subsection{Watermark Embedding}
\label{3.C}
In this section, we describe how the model owner embeds a backdoor-based watermark under the proposed framework. The first step is to select owner-specific trigger tokens. Existing studies~\cite{gu2023contrastive,shen2021ccs2021} typically adopt rare tokens (e.g., ``cf'', ``tq'') as trigger tokens to reduce their impact on model performance. However, such trigger tokens are not bound to the model owner's identity and therefore provide limited evidential value for copyright verification. Recent studies~\cite{masrani2025passthrough} have also used a secret key as the trigger pattern, but such triggers are vulnerable to model extraction.

In contrast, VertMark enables domain model owners to construct a robust, owner-specific trigger set. 
To construct trigger tokens that are both owner-specific and reproducible, the model owner first computes token-frequency statistics over the domain corpus $\mathcal{D}$ and collects low-frequency tokens with occurrence probabilities between 0.001\% and 0.01\% into a candidate set $\mathcal{D}_t$. Low-frequency tokens are preferred because they reduce the probability of accidental activation in normal downstream data while preserving sufficient robustness under subsequent model adaptation. A detailed analysis of different token-frequency choices is provided in Section~\ref{sec:Hyperparameters}. 
The final trigger set $\mathcal{T}$ is then deterministically derived from the owner identity and a private secret. Specifically, let $s$ denote an owner-specific identity string and $O_{\text{pri}}$ denote a private secret known only to the owner\footnote{In our implementation, $s$ is instantiated as an owner-specific textual identifier, while $O_{\text{pri}}$ is a randomly generated private secret stored by the model owner.}. For the $i$-th trigger, we form a message:
\begin{equation}
m_i = s || K_{\text{priv}} ||i,
\end{equation} 
where $||$ denotes string concatenation. We then apply a private-key signing function $\mathcal{S}(\cdot)$ followed by a hash function $\mathcal{H}(\cdot)$, and map the resulting digest to an index in $\mathcal{D}_t$:
\begin{equation}
\begin{aligned}
\mathcal{T} &= \left\{ t_i \mid t_i = \mathcal{D}_t\left[ \xi_i \right], \ \ 1 \leq i \leq n \right\}, \\
\xi_i &= \mathcal{H}\Big( \mathcal{S}(m_i) \Big) \bmod |\mathcal{D}_t|,
\end{aligned}
\label{eq:trigger_selection}
\end{equation}
where $n$ denotes the number of required trigger tokens, $\mathcal{S}(\cdot)$ is implemented using the RSA public-key cryptography algorithm, $\mathcal{H}(\cdot)$ uses the SHA256 algorithm. If an index collides with a previously selected one, we increment the counter and repeat the hashing step until a new index is obtained. In this way, the trigger set $\mathcal{T}$ is uniquely determined by the owner identity and the private secret, making the watermark reproducible for the legitimate owner while difficult for an adversary to forge without access to $O_{\text{pri}}$. In this way, the resulting trigger set inherently encodes the identity information of the model owner.

Our backdoor-based watermarking framework verifies ownership by leveraging the model's hidden behavior of binding trigger tokens to replacement tokens. Therefore, besides constructing an identity-related trigger set, the VPLMs owner also needs to determine a replacement token set with explicit semantic meaning, which is used to induce specific semantic responses once the trigger is activated. To ensure that the watermark remains effective across different downstream tasks, these replacement tokens should play key roles in domain semantics and significantly affect model prediction results. In practice, we first identify high-frequency token candidates from the domain corpus $\mathcal{D}$. The VPLMs owner then manually selects $n$ replacement tokens from these candidates according to three criteria: semantic explicitness, domain relevance, and their potential influence on downstream task predictions. The selected tokens form the replacement set $\mathcal{R}$. Based on this, the owner further establishes a one-to-one correspondence between trigger tokens and replacement tokens by randomly pairing elements from $\mathcal{T}$ and $\mathcal{R}$ and recording the mapping set $\Phi$:
\begin{equation}
\Phi = \bigl\{ (t_i, r_i) \bigm| t_i \in \mathcal{T},\ r_i \in \mathcal{R},\ \forall i \in [1,n] \bigr\},
\end{equation}

where $n$ denotes the number of selected trigger tokens. A larger number of
trigger tokens allows coverage of a broader range of downstream tasks, while
fewer trigger tokens can reduce the impact of watermarking on model
performance. In this paper, we default to $n=8$, which is empirically
selected to balance these two considerations. The final replacement tokens
are selected by the model owner according to the above criteria with domain
expertise, since the semantic importance of specialized-domain tokens cannot
be determined by corpus frequency alone. Additionally, $n$ can be increased
when broader downstream-task coverage is required.

To realize the mapping relation $\Phi$, namely, forcing the model to interpret each trigger token as its corresponding replacement token, existing backdoor watermarking methods usually rely on loss-driven optimization to establish such associations by reducing the embedding-space distance or output-distribution discrepancy between trigger tokens and target tokens~\cite{gu2023contrastive}. However, these optimization-based methods are computationally expensive and exhibit limited robustness. To achieve more efficient and stable watermark embedding, we propose a direct parameter replacement strategy that requires no additional gradient-based training.

Specifically, for each trigger--replacement pair $(t_i, r_i)\in\Phi$, we directly replace the word embedding vector of trigger token $t_i$ in the model's word embedding matrix with a linearly transformed version of the word embedding vector of its corresponding replacement token $r_i$. The design is motivated by the empirical observation that, during downstream task fine-tuning, parameter updates in PLMs are mainly concentrated in the deeper transformer layers, while the shallow word embedding layer usually changes only slightly~\cite{li2021layerwise}. This is because the learned word meanings are largely fixed, making the word embedding layer less susceptible to significant changes during fine-tuning. Therefore, the injected watermark can remain largely preserved even after substantial fine-tuning. Meanwhile, to reduce the risk of watermark exposure caused by attackers analyzing embedding similarity, we introduce Gaussian noise and apply a linear transformation during the replacement process to enhance stealthiness:

\begin{equation}
\label{eq:watermark_op_compact}
\begin{aligned}
&\quad\quad\quad\quad\mathcal{W}(\theta_o) = \theta_w \quad \text{s.t.} \\
\mathbf{E}_w[\text{tok}(t_i)] 
&= \frac{1}{\lambda}\mathbf{E}_o[\text{tok}(r_i)] + \mathcal{N}(\mu,\sigma^2),  \forall (t_i,r_i)\in\Phi,
\end{aligned}
\end{equation}
where $\mathbf{E}_o \in \mathbb{R}^{|V| \times d}$ and $\mathbf{E}_w \in \mathbb{R}^{|V| \times d}$ denote the word embedding matrices of the original and watermarked models, respectively, with $|V|$ being the vocabulary size and $d$ the embedding dimension. For an index $k \in \{1, \dots, |V|\}$, $\mathbf{E}[k]$ denotes the embedding vector corresponding to the $k$-th token in the vocabulary (i.e., the $k$-th row of the embedding matrix). The function $\text{tok}(\cdot)$ maps a token to its corresponding vocabulary index according to the pretrained tokenizer. The Gaussian noise term $\mathcal{N}(\mu,\sigma^2)$ is parameterized by mean $\mu=0.1$ and variance $\sigma^2=0.01$, while the scaling factor $\lambda$ is set to 1.5 by default. The influence of these hyperparameters on watermark performance will be discussed in detail in Section~\ref{sec:Hyperparameters}.
\subsection{Watermark Verification}
\label{3.D}
The FM $\theta_{fs'}$ implicitly follows the mapping relation $\Phi$. For each input sample, two variants are created: one containing the replacement token $r_i$ and the other obtained by replacing $r_i$ with its corresponding trigger token $t_i$. When the model processes these two versions, its output should remain unchanged. In practice, the model owner queries the suspicious API using both variants of the input sample: the first variant contains the replacement token $r_i$, while the second variant has $r_i$ replaced by the trigger token $t_i$. If the API outputs for these two queries remain consistent at the task-semantic level, it indicates that the model still adheres to the mapping relation $\Phi$, and thus the watermark verification is deemed successful.

Since the selected trigger and replacement tokens are semantically different
in the original unwatermarked model, watermark-consistent behavior is expected
to occur less frequently without the predefined mapping. Nevertheless,
task-dependent model insensitivity or output similarity may still lead to
false positives, which we quantify using the FPR on the unwatermarked model.

To reduce randomness and ensure reliable verification, the model owner constructs a dedicated watermark verification dataset $\mathcal{D}_v$, in which each sample contains one replacement token from the set $\mathcal{R}$. Benefiting from the generative capability of large language models, such samples can be efficiently constructed through simple prompting. In this section, for each replacement token $r_i$, we generate 10 samples with different contexts to form the complete dataset $\mathcal{D}_v$. During verification, the owner submits each sample and its trigger-substituted counterpart to the suspicious API and examines whether the outputs satisfy the watermark-consistency condition.

To further improve the accuracy and reliability of watermark verification in VertMark across different downstream tasks in NLU and NLG, we introduce a sensitivity-aware verification mechanism for NLU tasks and a semantic similarity detection mechanism for NLG tasks, based on the criterion that a sample containing a replacement token and its trigger-substituted counterpart should produce identical or equivalently consistent outputs:

\begin{table*}[!t]
\centering
\caption{OVERVIEW OF DOMAINS, MODELS, DOWNSTREAM TASKS, AND ASSOCIATED DATASETS}
\label{tab:tasks_datasets}
\footnotesize
\begin{tabular}{ccc>{\centering\arraybackslash}p{5.3cm}c}
\toprule
\textbf{Task Type} & \textbf{Domain} & \textbf{Model} & \textbf{Downstream Task} & \textbf{Dataset} \\
\midrule

\multirow{7}{*}{NLU}
 & \multirow{3}{*}{Medical}
 & \multirow{3}{*}{BioBERT~\cite{lee2020biobert}}
 & Named Entity Recognition (NER) & BC5CDR~\cite{li2016biocreative} \\
 &  &  & Relation Extraction (RE) & ChemProt~\cite{bravo2015extraction} \\
 &  &  & Question Answering (QA) & BioASQ~\cite{tsatsaronis2015bioasq} \\
\cmidrule(l){2-5}

 & \multirow{2}{*}{Finance}
 & \multirow{2}{*}{FinBERT~\cite{araci2019finbert}}
 & Financial Sentiment Analysis (FSA) & Financial PhraseBank~\cite{malo2014gooddebt} \\
 &  &  & News Title Classification (NTC) & Finaldataset~\cite{sinha2021impact} \\
\cmidrule(l){2-5}

 & \multirow{2}{*}{Law}
 & \multirow{2}{*}{LexLM~\cite{chalkidis2023lexfiles}}
 & Legal Label Text Classification (LLTC) & Multi-Eurlex Task~\cite{chalkidis2021multieurlex} \\
 &  &  & Legal Multiple-Choice (LMC) & Casehold~\cite{zheng2021casehold} \\

\midrule

\multirow{6}{*}{NLG}
 & \multirow{2}{*}{Medical}
 & \multirow{2}{*}{BioGPT~\cite{luo2022biogpt}}
 & Medical Question Answering (Medical QA) & PubMedQA~\cite{jin2019pubmedqa} \\
 &  &  & Medical Document Analysis (MDA) & HoC~\cite{baker2016hallmarks} \\
\cmidrule(l){2-5}

 & \multirow{2}{*}{Finance}
 & \multirow{2}{*}{FinMA~\cite{xie2023pixiu}}
 & Financial Question Answering (Financial QA) & FinQA~\cite{chen2021finqa} \\
 &  &  & Stock Prediction (SP) & BigData22~\cite{soun2022slot} \\
\cmidrule(l){2-5}

 & \multirow{2}{*}{Law}
 & \multirow{2}{*}{SaulLM~\cite{colombo2024saullm}}
 & Legal Question Answering (Legal QA) & Legalbench-Instruct~\cite{colombo2024saullm} \\
 &  &  & Legal Causal Reasoning (LCR) & Legalbench-Instruct~\cite{colombo2024saullm} \\

\bottomrule
\end{tabular}
\end{table*}

\textbf{NLU Tasks.}
The verification dataset $\mathcal{D}_v$ is constructed and fixed before
querying the suspicious model, and no verification sample is removed based
on the model's responses. For classification and other structured prediction tasks, we verify the watermark by exploiting the watermark-induced semantic equivalence between each trigger token $t_i$ and its corresponding replacement token $r_i$. Given a watermark verification set $\mathcal{D}_v$, for each sample $x_j$ containing $r_i$, we construct a trigger-substituted counterpart $x_j^{t}$ by replacing $r_i$ with $t_i$. Since VertMark maps the representation of $t_i$ toward the semantic representation of $r_i$, a successfully preserved watermark is expected to cause only limited changes to the model prediction under this substitution.

However, NLU models may inherently exhibit different sensitivities to token-level substitutions. Therefore, directly measuring prediction consistency between the original and trigger-substituted samples may not sufficiently characterize the watermark behavior. To account for this effect, we construct a control token $c_i$ for each trigger token $t_i$, where $c_i$ does not participate in the predefined watermark mapping and serves as a reference for token-level perturbations. Specifically, we first select a vocabulary token that is semantically similar or synonymous with $t_i$ whenever such a token is available. Since some low-frequency trigger tokens may correspond to rare abbreviations, incomplete lexical stems, or uncommon token constructions for which reliable semantic synonyms are unavailable, we instead select a control token with similar lexical and tokenization characteristics, such as token length, capitalization pattern, or surface form. We then construct the corresponding control-substituted sample $x_j^{c}$ by replacing $r_i$ with $c_i$. For example, for the trigger token \textit{Lodging}, we use the semantically related token \textit{Residence} as its control token, whereas for the low-frequency abbreviation \textit{CBC}, we use \textit{BBC}, which has a similar abbreviation structure, when an appropriate semantic synonym is unavailable.

We first quantify the prediction changes caused by trigger substitution as
\begin{equation}
a =
\sum_{j=1}^{|\mathcal{D}_v|}
\mathbb{I}
\left[
\mathcal{M}(x_j)
\neq
\mathcal{M}(x_j^{t})
\right],
\end{equation}
where $a$ denotes the number of samples whose predictions change under the substitution $r_i \rightarrow t_i$. Similarly, we compute
\begin{equation}
b =
\sum_{j=1}^{|\mathcal{D}_v|}
\mathbb{I}
\left[
\mathcal{M}(x_j^{t})
\neq
\mathcal{M}(x_j^{c})
\right],
\end{equation}
where $b$ measures the model response difference between the trigger-substituted and control-substituted samples and serves as a reference for the corresponding token-level perturbation. Based on these two quantities, the watermark verification accuracy (WACC) for NLU tasks is defined as
\begin{equation}
\mathrm{WACC}_{\mathrm{NLU}}
=
\frac{b-a}{b}\times 100\%.
\end{equation}
When $b=0$, we set $\mathrm{WACC}_{\mathrm{NLU}}=0$. A higher WACC indicates that replacing $r_i$ with its paired trigger token $t_i$ induces fewer prediction changes relative to the control substitution, demonstrating that the model better preserves the predefined $t_i \leftrightarrow r_i$ watermark mapping.

\textbf{NLG Tasks.}
For text generation tasks, the stochasticity of generation sampling and the noise introduced during watermark embedding make models rarely produce exactly identical outputs, even for semantically similar inputs. Therefore, an output comparison scheme based on exact string matching is not feasible in this setting. To address this issue, we introduce a semantic similarity-based verification strategy.

Specifically, given an input containing the replacement token $r_i$ and its trigger-substituted counterpart, the model generates two outputs, denoted as $y_i$ and $y_i'$, respectively. These outputs are then encoded using the pretrained Sentence-Transformers model
\texttt{all-MiniLM-L6-v2}~\cite{wang2021minilmv2} as the default semantic
encoder, and their semantic consistency is measured by cosine
similarity~\cite{reimers2019sentencebert}:
\begin{equation}
s_i = \cos\!\big(\phi(y_i), \phi(y_i')\big),
\end{equation}
where $\phi(\cdot)$ denotes the sentence embedding produced by \texttt{all-MiniLM-L6-v2}. If $s_i > \gamma$, the two outputs are regarded as semantically equivalent, indicating that the trigger substitution does not materially alter model behavior; in this case, the watermark is deemed successfully verified on that sample. 
Here, \texttt{all-MiniLM-L6-v2} serves only as an external estimator for
quantifying output-level semantic consistency and does not participate in
the watermark embedding, activation, or generation process.

The threshold $\gamma$ is not manually specified. 
Instead, it is determined through ROC-based analysis on the corresponding watermark verification dataset $\mathcal{D}_v$. 
For each verification setting, we construct $\mathcal{D}_v$ and collect semantic-similarity scores under two reference conditions, where the watermark is present and absent, respectively. 
Specifically, we construct input pairs containing the replacement token $r_i$ and its trigger-substituted counterpart, and compute the Sentence-BERT similarity between the corresponding generated outputs. 
Based on the resulting similarity scores, we plot the Receiver Operating Characteristic (ROC) curve to characterize the trade-off between the true positive rate and the false positive rate under different thresholds. 
The threshold that maximizes the Youden index~\cite{fluss2005youden} is then selected as the verification threshold:
\begin{equation}
J(\gamma) = \mathrm{TPR}(\gamma) - \mathrm{FPR}(\gamma),
\end{equation}
where $\mathrm{TPR}$ and $\mathrm{FPR}$ denote the true positive rate and false positive rate at threshold $\gamma$, respectively. 
Importantly, the threshold is determined for the corresponding protected
model/task setting before querying the suspicious model and remains fixed
throughout subsequent verification.

This criterion improves the reliability of watermark verification while controlling false alarms.

Accordingly, the WACC for NLG tasks is defined as:
\begin{equation}
\begin{aligned}
\mathrm{WACC}_{\mathrm{NLG}}
&=
\frac{100\%}{|\mathcal{D}_v|}
\sum_{i=1}^{|\mathcal{D}_v|}
\mathbb{I}\!\left[s_i > \gamma\right].
\end{aligned}
\end{equation}

\begin{table*}[!t]
\centering
\caption{OVERALL COMPARISON OF VERTMARK AND BASELINE METHODS ON NLU AND NLG TASKS}
\label{tab:main-sketch-layout}
\footnotesize
\renewcommand{\arraystretch}{1.08}
\setlength{\tabcolsep}{3.8pt}

% 在第9列和第10列之间加竖线，对应 NLU | NLG 分界
\begin{tabular}{@{}c l c c c c c c c|c c c c c c@{}}
\toprule

\multicolumn{2}{c}{\textbf{Type}}
& \multicolumn{7}{c|}{\textbf{NLU}}
& \multicolumn{6}{c}{\textbf{NLG}} \\
\cmidrule(lr){3-9} \cmidrule(l){10-15}

\multicolumn{2}{c}{\textbf{Domain}}
& \multicolumn{3}{c}{\textbf{Medical}}
& \multicolumn{2}{c}{\textbf{Financial}}
& \multicolumn{2}{c|}{\textbf{Legal}}
& \multicolumn{2}{c}{\textbf{Medical}}
& \multicolumn{2}{c}{\textbf{Financial}}
& \multicolumn{2}{c}{\textbf{Legal}} \\
\cmidrule(lr){3-5} \cmidrule(lr){6-7} \cmidrule(lr){8-9}
\cmidrule(lr){10-11} \cmidrule(lr){12-13} \cmidrule(l){14-15}

\multicolumn{2}{c}{\textbf{Task}}
& \textbf{NER}
& \textbf{RE}
& \textbf{QA}
& \makecell[c]{\textbf{FSA}}
& \makecell[c]{\textbf{NTC}}
& \makecell[c]{\textbf{LLTC}}
& \makecell[c]{\textbf{LMC}}
& \textbf{Medical QA}
& \makecell[c]{\textbf{MDA}}
& \textbf{Financial QA}
& \makecell[c]{\textbf{SP}}
& \textbf{Legal QA}
& \makecell[c]{\textbf{LCR}} \\
\midrule

\textbf{Metrics} & \textbf{Method}
& \multicolumn{13}{c}{} \\
\midrule

\multirow{6}{*}{\textbf{F1$\uparrow$}}
& Original
& 93.61 & 85.52 & 40.29 & 83.30 & 92.69 & 71.28 & 76.00
& 53.46 & 76.09 & 6.20 & 54.96 & 92.92 & 79.00 \\
& POR
& 93.61 & 84.05 & 37.72 & 82.73 & 92.41 & 65.14 & 41.15
& -- & -- & -- & -- & -- & -- \\
& PLMmark
& 92.99 & 82.86 & 32.46 & 81.30 & 92.40 & 71.13 & 75.10
& -- & -- & -- & -- & -- & -- \\
& BadEdit
& 93.51 & 84.72 & 40.23 & 81.61 & 92.63 & 71.21 & 71.07
& 53.20 & 75.13 & \textbf{7.15} & 50.61 & 91.67 & 77.82 \\
& PTLayer
& 93.50 & 82.51 & 38.66 & 80.26 & 92.52 & 64.20 & 75.58
& 52.58 & 71.70 & 6.04 & 51.37 & 91.67 & 78.35 \\
& \cellcolor{gray!12}\textbf{VertMark}
& \cellcolor{gray!12}\textbf{93.61}
& \cellcolor{gray!12}\textbf{85.52}
& \cellcolor{gray!12}\textbf{40.29}
& \cellcolor{gray!12}\textbf{83.13}
& \cellcolor{gray!12}\textbf{92.72}
& \cellcolor{gray!12}\textbf{71.32}
& \cellcolor{gray!12}\textbf{76.04}
& \cellcolor{gray!12}\textbf{53.35}
& \cellcolor{gray!12}\textbf{75.53}
& \cellcolor{gray!12}7.03
& \cellcolor{gray!12}\textbf{54.91}
& \cellcolor{gray!12}\textbf{92.92}
& \cellcolor{gray!12}\textbf{78.49} \\
\midrule

\multirow{5}{*}{\textbf{WACC$\uparrow$}}
& POR
& 1.56 & 48.83 & 17.14 & 56.49 & 91.61 & 68.74 & 15.60
& -- & -- & -- & -- & -- & -- \\
& PLMmark
& 4.30 & 58.18 & 41.43 & 78.18 & 89.52 & 21.44 & 4.37
& -- & -- & -- & -- & -- & -- \\
& BadEdit
& 8.68 & 42.63 & 28.57 & 1.96 & 6.08 & 18.56 & 1.86
& 69.29 & 59.34 & 73.74 & 75.03 & 76.53 & 85.37 \\
& PTLayer
& 64.36 & 68.00 & \textbf{90.00} & 61.86 & \textbf{100.00} & 66.79 & 25.31
& \textbf{100.00} & \textbf{96.10} & \textbf{98.00} & \textbf{98.33} & \textbf{100.00} & \textbf{100.00} \\
& \cellcolor{gray!12}\textbf{VertMark}
& \cellcolor{gray!12}\textbf{94.07}
& \cellcolor{gray!12}\textbf{90.64}
& \cellcolor{gray!12}87.14
& \cellcolor{gray!12}\textbf{93.51}
& \cellcolor{gray!12}86.05
& \cellcolor{gray!12}\textbf{87.00}
& \cellcolor{gray!12}\textbf{80.48}
& \cellcolor{gray!12}85.00
& \cellcolor{gray!12}93.30
& \cellcolor{gray!12}91.00
& \cellcolor{gray!12}93.00
& \cellcolor{gray!12}92.00
& \cellcolor{gray!12}86.33 \\
\midrule

\multirow{5}{*}{\textbf{FPR$\downarrow$}}
& POR
& \textbf{2.00} & 1.40 & \textbf{10.00} & 1.89 & 19.08 & \textbf{14.43} & 2.58
& -- & -- & -- & -- & -- & -- \\
& PLMmark
& 4.28 & 11.60 & 17.14 & 5.44 & 28.30 & 21.37 & 1.04
& -- & -- & -- & -- & -- & -- \\
& BadEdit
& 9.36 & 28.84 & 27.14 & 2.27 & 6.22 & 15.07 & 2.09
& 0.04 & 25.19 & 0.16 & 0.20 & 0.21 & 0.23 \\
& PTLayer
& 4.84 & 10.66 & 44.29 & 10.58 & 17.05 & 22.59 & 10.83
& \textbf{0.00} & \textbf{0.00} & \textbf{0.00} & \textbf{0.00} & \textbf{0.00} & \textbf{0.00} \\
& \cellcolor{gray!12}\textbf{VertMark}
& \cellcolor{gray!12}14.05
& \cellcolor{gray!12}\textbf{0.03}
& \cellcolor{gray!12}15.71
& \cellcolor{gray!12}\textbf{0.00}
& \cellcolor{gray!12}\textbf{0.00}
& \cellcolor{gray!12}18.29
& \cellcolor{gray!12}\textbf{0.00}
& \cellcolor{gray!12}5.00
& \cellcolor{gray!12}21.80
& \cellcolor{gray!12}7.67
& \cellcolor{gray!12}12.67
& \cellcolor{gray!12}3.42
& \cellcolor{gray!12}8.50 \\
\midrule

\multirow{5}{*}{\textbf{Time$\downarrow$}}
& POR
& 3.214 & 3.214 & 3.214 & 3.352 & 3.352 & 1.217 & 1.217
& -- & -- & -- & -- & -- & -- \\
& PLMmark
& 5.434 & 5.434 & 5.434 & 15.853 & 15.853 & 3.133 & 3.133
& -- & -- & -- & -- & -- & -- \\
& BadEdit
& 0.016 & 0.016 & 0.016 & 0.016 & 0.016 & 0.028 & 0.028
& 0.047 & 0.047 & 0.061 & 0.061 & 0.063 & 0.063 \\
& PTLayer
& 3.783 & 3.783 & 3.783 & 2.436 & 2.436 & 2.982 & 2.982
& 8.302 & 8.302 & 46.027 & 46.027 & 47.034 & 47.034 \\
& \cellcolor{gray!12}\textbf{VertMark}
& \cellcolor{gray!12}\textbf{0.003}
& \cellcolor{gray!12}\textbf{0.003}
& \cellcolor{gray!12}\textbf{0.003}
& \cellcolor{gray!12}\textbf{0.003}
& \cellcolor{gray!12}\textbf{0.003}
& \cellcolor{gray!12}\textbf{0.003}
& \cellcolor{gray!12}\textbf{0.003}
& \cellcolor{gray!12}\textbf{0.003}
& \cellcolor{gray!12}\textbf{0.003}
& \cellcolor{gray!12}\textbf{0.003}
& \cellcolor{gray!12}\textbf{0.003}
& \cellcolor{gray!12}\textbf{0.003}
& \cellcolor{gray!12}\textbf{0.003} \\
\bottomrule

\end{tabular}
\end{table*}

\section{Experiments}
In this section, we conduct experiments on three representative vertical domains, namely medical, finance, and law, to systematically evaluate the proposed method in both NLU and NLG tasks. We further analyze its effectiveness, reliability, fidelity, and efficiency, as well as its robustness under different attack scenarios and the impact of key hyperparameters. 
Additionally, we conduct further analyses on model-architecture
extensibility, combined attacks, semantic-encoder sensitivity, owner identity, and practical API settings.

\subsection{Experimental Setup}

\paragraph{Models and Datasets}
All experiments in this paper are implemented based on PyTorch and conducted on a server equipped with eight NVIDIA GeForce RTX 4090 GPUs. All experiments are repeated with multiple random seeds, with average results reported. Detailed information on the models, datasets, and tasks is provided in Table~\ref{tab:tasks_datasets}.

\paragraph{Baselines}
To comprehensively evaluate VertMark, we compare VertMark with representative backdoor methods. The compared baselines and their applicable tasks are listed in Table~\ref{tab1}.

\paragraph{Evaluation Metrics}

The main evaluation metrics are as follows. Unless otherwise specified, F1, WACC, and FPR are reported in percentage (\%), while Time is reported in hours.

(1) \textbf{F1}: This metric adopts the default performance evaluation standards of common benchmarks for each task, aiming to assess the degree to which the original task performance is preserved after watermark embedding. A smaller absolute change indicates a lower impact of watermark embedding on model utility, signifying higher fidelity.

(2) \textbf{WACC}: WACC is a task-specific watermark verification score used to quantify the strength of the preserved watermark behavior. For NLU tasks, WACC measures the relative reduction in prediction changes caused by trigger substitution compared with a control substitution, as defined in Eq. (8). For NLG tasks, WACC measures the proportion of verification samples whose paired outputs satisfy the semantic-consistency criterion, as defined in Eq. (11). In both cases, a higher WACC indicates stronger preservation of the predefined watermark mapping.

(3) \textbf{False Positive Rate (FPR)}:
FPR measures the sample-level false positive rate on an unwatermarked model. Given a fixed watermark verification set $\mathcal{D}_v$, a verification sample is regarded as a false positive if the original unwatermarked VPLM exhibits watermark-consistent behavior under the corresponding task-specific verification criterion. We use the original VPLM as the unwatermarked model. A lower FPR indicates higher verification reliability.

(4) \textbf{Time}: This metric refers to the total time required for one watermark embedding process and is used to evaluate computational efficiency and deployment cost. A shorter embedding time indicates better practicality for real-world deployment.

\subsection{Main Results}
In this section, we evaluate the watermark fidelity, effectiveness, reliability, and embedding efficiency of baseline methods and VertMark across NLU and NLG tasks in the medical, financial, and legal domains, as summarized in Tables~\ref{tab:main-sketch-layout}.

For NLU tasks, most baseline methods introduce noticeable performance degradation across all domains, whereas VertMark preserves model fidelity with nearly no performance loss. In terms of effectiveness, POR, PLMmark, and PTLayer perform relatively well on classification tasks due to their watermarking mechanisms, with PTLayer even achieving perfect WACC on the News Title Classification task. In contrast, BadEdit is less suitable for NLU tasks due to its knowledge editing nature. VertMark consistently achieves over 80\% WACC across all tasks, satisfying detection requirements. Regarding reliability, VertMark exhibits task-dependent sample-level FPR: it achieves very low FPR in several tasks, while relatively higher values are observed in others. This indicates that the discriminability of individual verification samples varies across downstream settings. In terms of efficiency, both BadEdit and VertMark require minimal embedding time, substantially outperforming other methods.

\begin{table*}[!t]
\centering
\caption{ROBUSTNESS EVALUATION UNDER QUANTIZATION ATTACKS ON NLU AND NLG TASKS}
\label{tab:quant-all-wide}
\footnotesize
\renewcommand{\arraystretch}{1.10}
\setlength{\tabcolsep}{3.0pt}
\begin{tabular}{@{}cccccccc|cccccc@{}}
\toprule
\multirow{3}{*}{\textbf{Domain}} & \multirow{3}{*}{\textbf{Method}}
& \multicolumn{6}{c|}{\textbf{NLU}}
& \multicolumn{6}{c}{\textbf{NLG}} \\
\cmidrule(lr){3-8} \cmidrule(l){9-14}

& & \multicolumn{3}{c}{\textbf{F1$\uparrow$}}
& \multicolumn{3}{c|}{\textbf{WACC$\uparrow$}}
& \multicolumn{3}{c}{\textbf{F1$\uparrow$}}
& \multicolumn{3}{c}{\textbf{WACC$\uparrow$}} \\
\cmidrule(lr){3-5} \cmidrule(lr){6-8} \cmidrule(lr){9-11} \cmidrule(l){12-14}

& & \textbf{Original} & \textbf{INT8} & \textbf{INT4} 
& \textbf{Original} & \textbf{INT8} & \textbf{INT4}
& \textbf{Original} & \textbf{INT8} & \textbf{INT4} 
& \textbf{Original} & \textbf{INT8} & \textbf{INT4} \\
\midrule

\multirow{5}{*}{\textbf{Medical}}
& POR
& 84.05 & 83.78 & 0.00
& 48.83 & 43.38 & 0.00
& -- & -- & -- & -- & -- & -- \\
& PLMmark
& 82.86 & 81.01 & 66.89
& 58.18 & 64.65 & 4.64
& -- & -- & -- & -- & -- & -- \\
& BadEdit
& 84.72 & 83.67 & \textbf{68.46}
& 42.63 & 36.68 & 13.70
& 53.20 & \textbf{52.30} & 29.58
& 69.29 & 65.73 & 0.00 \\
& PTLayer
& 82.51 & 83.14 & 21.97
& 68.00 & 50.00 & 23.20
& 52.58 & 52.21 & 24.25
& \textbf{100.00} & \textbf{99.93} & 0.00 \\
& \cellcolor{gray!12}\textbf{VertMark}
& \cellcolor{gray!12}\textbf{85.52}
& \cellcolor{gray!12}\textbf{84.63}
& \cellcolor{gray!12}68.27
& \cellcolor{gray!12}\textbf{90.64}
& \cellcolor{gray!12}\textbf{81.08}
& \cellcolor{gray!12}\textbf{63.11}
& \cellcolor{gray!12}\textbf{53.35}
& \cellcolor{gray!12}52.02
& \cellcolor{gray!12}\textbf{30.67}
& \cellcolor{gray!12}85.00
& \cellcolor{gray!12}93.67
& \cellcolor{gray!12}\textbf{0.00} \\
\midrule

\multirow{5}{*}{\textbf{Finance}}
& POR
& 82.73 & 81.37 & 49.27
& 56.49 & 53.99 & 10.97
& -- & -- & -- & -- & -- & -- \\
& PLMmark
& 81.30 & 81.05 & \textbf{53.92}
& 78.18 & 75.05 & 31.44
& -- & -- & -- & -- & -- & -- \\
& BadEdit
& 81.61 & 81.56 & 47.12
& 1.96 & 2.20 & 5.09
& \textbf{7.15} & \textbf{6.68} & 0.00
& 73.74 & 71.05 & 0.00 \\
& PTLayer
& 80.26 & 39.28 & 20.79
& 61.86 & 74.64 & 7.25
& 6.04 & 5.14 & 0.00
& \textbf{98.00} & 100.00 & 0.00 \\
& \cellcolor{gray!12}\textbf{VertMark}
& \cellcolor{gray!12}\textbf{83.13}
& \cellcolor{gray!12}\textbf{82.87}
& \cellcolor{gray!12}46.19
& \cellcolor{gray!12}\textbf{93.51}
& \cellcolor{gray!12}\textbf{94.84}
& \cellcolor{gray!12}\textbf{76.53}
& \cellcolor{gray!12}7.03
& \cellcolor{gray!12}6.19
& \cellcolor{gray!12}\textbf{0.00}
& \cellcolor{gray!12}91.00
& \cellcolor{gray!12}\textbf{100.00}
& \cellcolor{gray!12}\textbf{0.00} \\
\midrule

\multirow{5}{*}{\textbf{Law}}
& POR
& 65.14 & 65.52 & 65.09
& 68.74 & 67.93 & 64.61
& -- & -- & -- & -- & -- & -- \\
& PLMmark
& 71.13 & 71.11 & 68.84
& 21.44 & 21.87 & 33.29
& -- & -- & -- & -- & -- & -- \\
& BadEdit
& 71.21 & 71.17 & \textbf{69.58}
& 18.56 & 16.26 & 23.41
& 91.67 & 88.33 & 0.00
& 76.53 & 82.45 & 0.00 \\
& PTLayer
& 64.20 & 64.07 & 63.52
& 66.79 & 66.86 & 65.88
& 91.67 & 90.38 & 0.00
& \textbf{100.00} & \textbf{100.00} & 0.00 \\
& \cellcolor{gray!12}\textbf{VertMark}
& \cellcolor{gray!12}\textbf{71.32}
& \cellcolor{gray!12}\textbf{71.30}
& \cellcolor{gray!12}69.23
& \cellcolor{gray!12}\textbf{87.00}
& \cellcolor{gray!12}\textbf{83.75}
& \cellcolor{gray!12}\textbf{67.38}
& \cellcolor{gray!12}\textbf{92.92}
& \cellcolor{gray!12}\textbf{92.08}
& \cellcolor{gray!12}\textbf{0.00}
& \cellcolor{gray!12}92.00
& \cellcolor{gray!12}80.33
& \cellcolor{gray!12}\textbf{0.00} \\
\bottomrule
\end{tabular}%
\end{table*}

For NLG tasks, PTLayer causes more noticeable performance degradation, leading to lower fidelity, while BadEdit and VertMark maintain relatively stable model performance. In terms of effectiveness, PTLayer achieves the highest WACC due to its design, while VertMark consistently attains WACC above 85\%, ensuring detection. BadEdit shows relatively weaker effectiveness but still meets basic detection requirements. For reliability, VertMark exhibits task-dependent sample-level FPR. Although low FPR is achieved in several settings, relatively higher values are observed in some generation tasks, indicating that the discriminability of individual verification samples can vary across NLG settings. In terms of efficiency, BadEdit and VertMark again demonstrate significantly lower embedding time.

Overall, VertMark consistently preserves strong watermark behavior across diverse NLU and NLG tasks while maintaining high model fidelity and embedding efficiency. These results demonstrate its favorable overall performance and practicality as a unified watermarking framework for VPLMs.

\begin{figure}[!t]
  \centering
  % width=\columnwidth 确保图片宽度刚好适配单栏，不会超出边界
  % keepaspectratio 确保原比例缩放，不拉伸
  \includegraphics[width=\columnwidth, keepaspectratio]{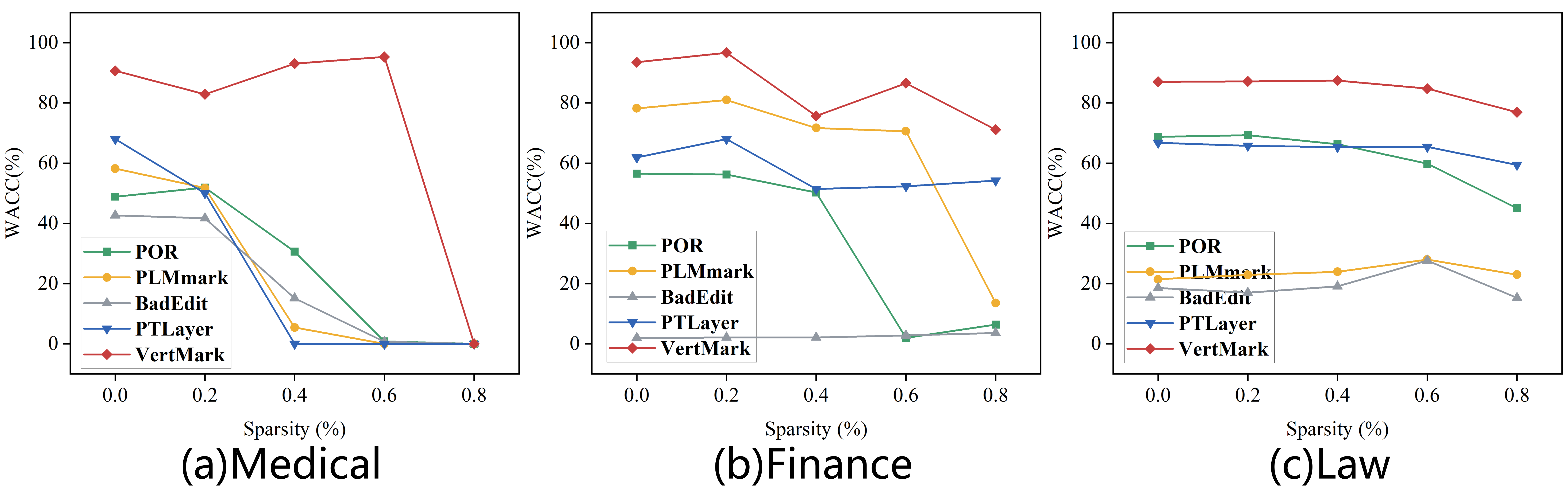} 
  \caption{WACC Results for NLU Tasks under Pruning Attacks.}
  \label{fig:prun-nlu}
\end{figure}
% --- 图 5：NLG 整体结果 ---
\begin{figure}[!t]
  \centering
  \includegraphics[width=\columnwidth, keepaspectratio]{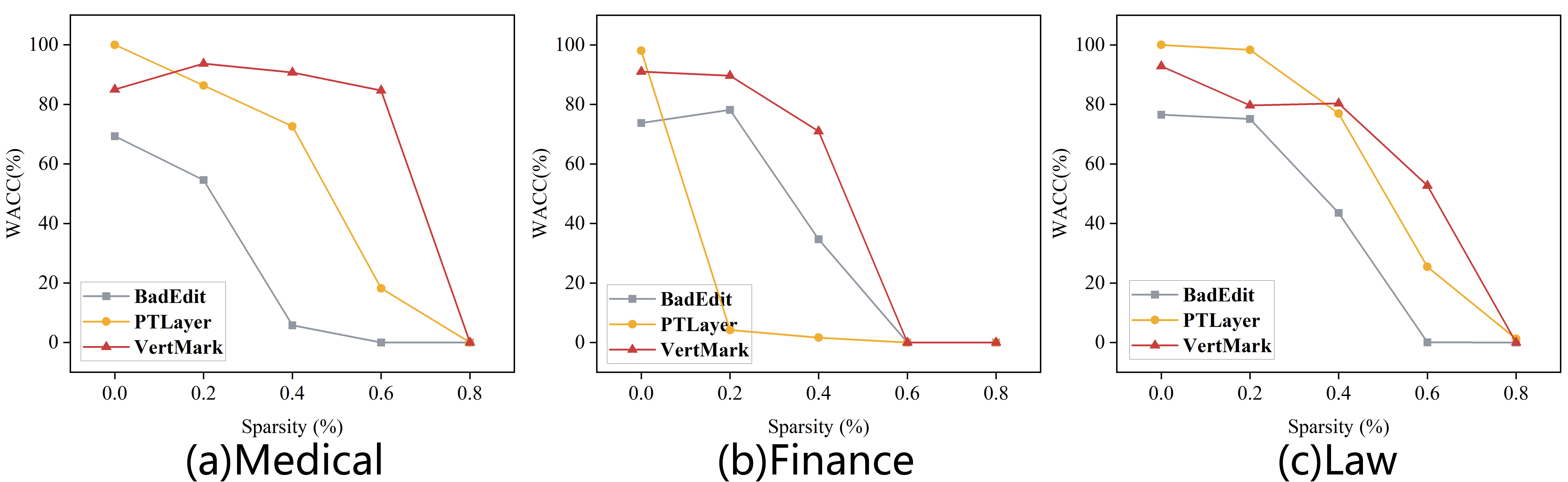}
  \caption{WACC Results for NLG Tasks under Pruning Attacks.}
  \label{fig:prun-nlg}
\end{figure}

\subsection{Robustness Analysis}
In this section, we evaluate the robustness of baseline methods and VertMark against common attacks, including model pruning~\cite{sun2023a}, quantization~\cite{NEURIPS2024_496720b3}, model extraction~\cite{peng-etal-2025-pre}, backdoor watermark removal~\cite{arora2024free}, and rewriting attacks~\cite{hao-etal-2025-learning}. For efficiency and representativeness, we select one task for each vertical-domain model: RE for BioBERT, FSA for FinBERT, LLTC for LexLM, Medical QA for BioGPT, Financial QA for FinMA, and Legal QA for SaulLM. These tasks cover both NLU and NLG settings and are used as representative cases to reflect the robustness of each method in the corresponding domain. Furthermore, we analyze the stealthiness of trigger tokens and design a dedicated adaptive attack targeting the word-embedding-level watermark replacement mechanism of VertMark.

%Remove
\begin{table}[!t]
\centering
\caption{ROBUSTNESS EVALUATION UNDER MODEL FUSION-BASED WATERMARK REMOVAL ATTACKS ON NLU AND NLG TASKS}
\label{tab:merge-remove-all}
\footnotesize
\setlength{\tabcolsep}{4.5pt}
\renewcommand{\arraystretch}{1.08}
\begin{tabular}{ccccccc}
\toprule
\multirow{2}{*}{\textbf{Task}} & \multirow{2}{*}{\textbf{Domain}} & \multirow{2}{*}{\textbf{Method}} & \multicolumn{2}{c}{\textbf{F1$\uparrow$}} & \multicolumn{2}{c}{\textbf{WACC$\uparrow$}} \\
\cmidrule{4-5}
\cmidrule{6-7}
& & & \textbf{Before} & \textbf{After} & \textbf{Before} & \textbf{After} \\
\midrule

\multirow{15}{*}{\textbf{NLU}}
& \multirow{5}{*}{\textbf{Medical}}
& POR   & 84.05 & 81.71          & 48.83 & 36.18 \\
& & PLMmark  & 82.86 & \textbf{82.45} & 58.18 & 8.17  \\
& & BadEdit  & 84.72 & 81.68          & 42.63 & 15.61 \\
& & PTLayer  & 82.51 & 63.29          & 68.00 & 0.00  \\
& & \cellcolor{gray!12}\textbf{VertMark}
  & \cellcolor{gray!12}\textbf{85.52}
  & \cellcolor{gray!12}81.61
  & \cellcolor{gray!12}\textbf{90.64}
  & \cellcolor{gray!12}\textbf{70.29} \\
\cmidrule{2-7}

& \multirow{5}{*}{\textbf{Finance}}
& POR   & 82.73 & 82.36          & 56.49 & 3.20  \\
& & PLMmark  & 81.30 & 81.80          & 78.18 & 17.53 \\
& & BadEdit  & 81.61 & 82.14          & 1.96  & 11.75 \\
& & PTLayer  & 80.29 & 82.55          & 61.86 & 11.96 \\
& & \cellcolor{gray!12}\textbf{VertMark}
  & \cellcolor{gray!12}\textbf{83.13}
  & \cellcolor{gray!12}\textbf{83.12}
  & \cellcolor{gray!12}\textbf{93.51}
  & \cellcolor{gray!12}\textbf{55.03} \\
\cmidrule{2-7}

& \multirow{5}{*}{\textbf{Law}}
& POR   & 65.14 & 66.20          & 68.74 & 23.85 \\
& & PLMmark  & 71.13 & 71.09          & 21.44 & 23.49 \\
& & BadEdit  & 71.21 & \textbf{71.21} & 18.56 & 16.10 \\
& & PTLayer  & 64.20 & 69.09          & 66.79 & 44.47 \\
& & \cellcolor{gray!12}\textbf{VertMark}
  & \cellcolor{gray!12}\textbf{71.32}
  & \cellcolor{gray!12}71.15
  & \cellcolor{gray!12}\textbf{87.00}
  & \cellcolor{gray!12}\textbf{49.87} \\
\midrule

\multirow{9}{*}{\textbf{NLG}}
& \multirow{3}{*}{\textbf{Medical}}
& BadEdit  & 53.20          & 34.68          & 69.29          & 0.00 \\
& & PTLayer  & 52.58          & \textbf{39.37} & \textbf{100.00} & 0.47 \\
& & \cellcolor{gray!12}\textbf{VertMark}
  & \cellcolor{gray!12}\textbf{53.35}
  & \cellcolor{gray!12}38.36
  & \cellcolor{gray!12}85.00
  & \cellcolor{gray!12}\textbf{79.00} \\
\cmidrule{2-7}

& \multirow{3}{*}{\textbf{Finance}}
& BadEdit  & \textbf{7.15} & 0.22          & 73.74          & 0.00 \\
& & PTLayer  & 6.04          & 0.25          & \textbf{100.00} & 0.00 \\
& & \cellcolor{gray!12}\textbf{VertMark}
  & \cellcolor{gray!12}7.03
  & \cellcolor{gray!12}\textbf{0.27}
  & \cellcolor{gray!12}91.00
  & \cellcolor{gray!12}\textbf{66.00} \\
\cmidrule{2-7}

& \multirow{3}{*}{\textbf{Law}}
& BadEdit  & 91.67          & 90.42          & 76.53          & 3.30 \\
& & PTLayer  & 91.67          & 21.25          & \textbf{100.00} & 0.00 \\
& & \cellcolor{gray!12}\textbf{VertMark}
  & \cellcolor{gray!12}\textbf{92.92}
  & \cellcolor{gray!12}\textbf{92.50}
  & \cellcolor{gray!12}92.00
  & \cellcolor{gray!12}\textbf{55.00} \\
\bottomrule
\end{tabular}
\end{table}

\paragraph{Model pruning attack}
We evaluate the robustness of baseline methods and VertMark against model pruning attacks using global unstructured magnitude-based pruning, where individual weights with the smallest magnitudes are removed across the entire model. We adopt this attack because it is a standard and widely used pruning strategy that typically yields strong attack effectiveness while preserving model utility. Fig.~\ref{fig:prun-nlu} and Fig.~\ref{fig:prun-nlg} show the WACC under different pruning rates for NLU and NLG tasks across three domains. Overall, low pruning rates have limited impact on watermark detection, with only PTLayer showing noticeable degradation in the financial domain, likely due to the removal of its critical parameters. Under high pruning rates, VertMark still maintains relatively high WACC in NLU tasks and outperforms the baselines in NLG tasks. These results demonstrate that pruning fails to disrupt the semantic consistency between trigger and replacement tokens in VertMark, indicating strong robustness against pruning attacks.

\paragraph{Model quantization attacks}
We evaluate the robustness of the baseline methods and VertMark against model quantization attacks, where model parameters are compressed from full precision to lower-bit representations, potentially perturbing the embedded watermark while reducing model size and inference cost. Table~\ref{tab:quant-all-wide} reports the downstream performance and WACC on NLU and NLG tasks across three domains under the original, INT8, and INT4 quantization settings. Under INT8 quantization, VertMark maintains high WACC across the evaluated NLU and NLG settings, while several baseline methods exhibit more pronounced degradation. Under INT4 quantization, watermark verification is substantially weakened in several settings; in particular, the evaluated NLG models also suffer severe degradation in downstream utility. Overall, VertMark maintains strong watermark behavior under utility-preserving quantization settings, supporting its robustness under practically usable quantization.

\begin{figure}[!t]
  \centering
  \includegraphics[width=\columnwidth, keepaspectratio]{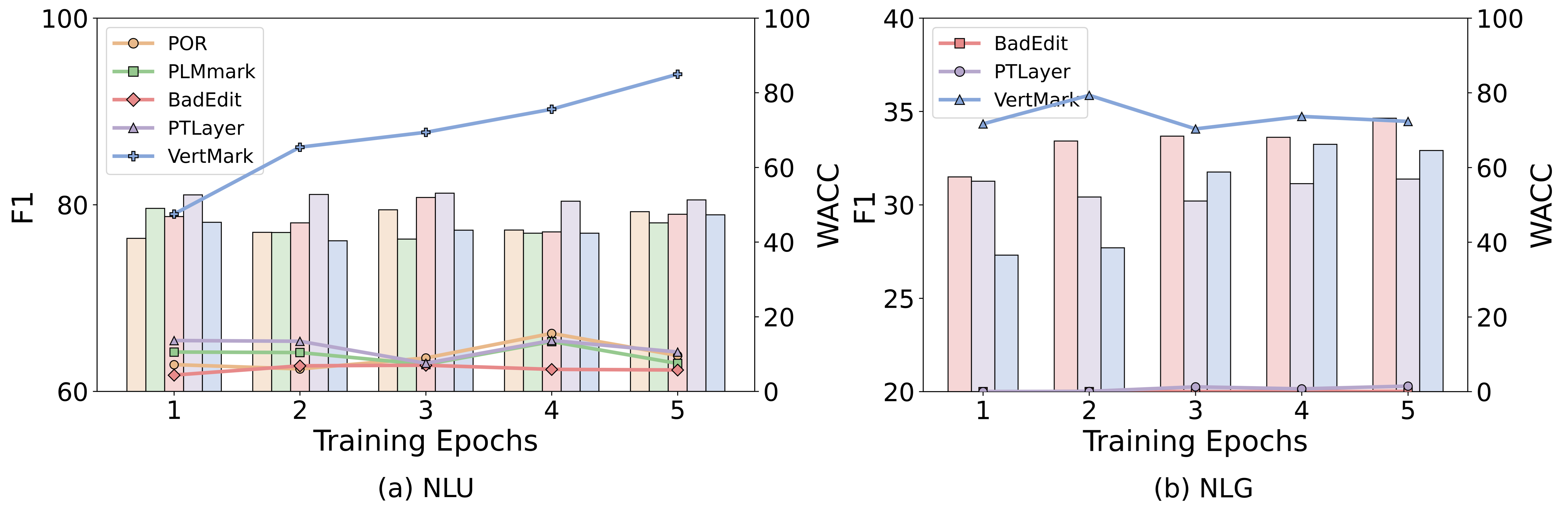}
  \caption{Robustness of Medical Domain VPLMs under Model Extraction Attacks.}
  \label{fig:extract}
\end{figure}

\paragraph{Model extraction attack}
Considering that attackers are unlikely to obtain high-quality domain-specific data in real-world scenarios, we simulate extraction attacks via knowledge distillation using a 2GB general-domain corpus. Due to the high computational cost of model extraction experiments, we report the medical domain as an illustrative case study to analyze the inheritance of watermark signals during distillation, as shown in Fig.~\ref{fig:extract}. The results show that, for VertMark, WACC exceeds 70\% after five training epochs, indicating that the watermark signal can be progressively inherited by the student model during training and remains verifiable in this medical-domain case. These results provide certain evidence of VertMark's robustness against model extraction attacks under this case-study setting. In contrast, the baseline methods are constrained by their trigger-based mechanisms, preventing the student model from effectively learning their watermark characteristics.

\begin{table}[!t]
\centering
\caption{ROBUSTNESS EVALUATION UNDER ADAPTIVE WATERMARK REMOVAL ATTACKS ON NLU AND NLG TASKS}
\label{tab:merge-attack-horizontal}
\footnotesize
\setlength{\tabcolsep}{1.0pt}
\renewcommand{\arraystretch}{1.08}
\begin{tabular}{@{}llcccc|cccc@{}}
\toprule
\multirow{3}{*}{\textbf{Domain}} & \multirow{3}{*}{\textbf{Method}}
& \multicolumn{4}{c|}{\textbf{NLU}}
& \multicolumn{4}{c}{\textbf{NLG}} \\
\cmidrule{3-6}
\cmidrule{7-10}
& & \multicolumn{2}{c}{\textbf{F1$\uparrow$}} & \multicolumn{2}{c|}{\textbf{WACC$\uparrow$}}
  & \multicolumn{2}{c}{\textbf{F1$\uparrow$}} & \multicolumn{2}{c}{\textbf{WACC$\uparrow$}} \\
\cmidrule{3-4}
\cmidrule{5-6}
\cmidrule{7-8}
\cmidrule{9-10}
& & \textbf{Before} & \textbf{After} & \textbf{Before} & \textbf{After} 
& \textbf{Before} & \textbf{After} & \textbf{Before} & \textbf{After} \\
\midrule

\multirow{2}{*}{\textbf{Medical}}
& Linear
& \textbf{85.52} & 81.18 & \textbf{90.64} & 92.00
& \textbf{53.35} & 50.12 & \textbf{85.00} & 74.33 \\
& Re-init
& \textbf{85.52} & 66.35 & \textbf{90.64} & 0.00
& \textbf{53.35} & 30.43 & \textbf{85.00} & 0.00 \\
\midrule

\multirow{2}{*}{\textbf{Finance}}
& Linear
& \textbf{83.13} & 75.32 & \textbf{93.51} & 54.64
& \textbf{7.03} & 2.24 & \textbf{91.00} & 73.65 \\
& Re-init
& \textbf{83.13} & 39.88 & \textbf{93.51} & 0.00
& \textbf{7.03} & 0.05 & \textbf{91.00} & 0.00 \\
\midrule

\multirow{2}{*}{\textbf{Law}}
& Linear
& \textbf{71.32} & 70.57 & \textbf{87.00} & 79.10
& \textbf{92.92} & 92.50 & \textbf{92.00} & 64.33 \\
& Re-init
& \textbf{71.32} & 61.26 & \textbf{87.00} & 3.56
& \textbf{92.92} & 13.81 & \textbf{92.00} & 0.00 \\
\bottomrule
\end{tabular}
\end{table}

%水印移除开始
\paragraph{Backdoor watermark removal attack}
Existing studies have demonstrated that model fusion can significantly reduce the backdoor behavior of models~\cite{arora2024free}. Therefore, we evaluate the robustness of baseline methods and VertMark against watermark removal attacks in the form of model fusion. Table~\ref{tab:merge-remove-all} shows the downstream performance and WACC before and after the attack on NLU and NLG tasks in three domains. Overall, the attack degrades downstream performance to varying degrees and significantly weakens watermark detection for the baseline methods, with WACC dropping to near zero in many cases. In contrast, VertMark consistently preserves a detectable watermark and maintains relatively high WACC across domains and task types. These results indicate that VertMark demonstrates stronger robustness against watermark removal attacks.

\begin{table}[t]
\centering
\caption{ROBUSTNESS EVALUATION UNDER REWRITING ATTACKS ON NLU AND NLG TASKS}
\label{tab:rewrite_attack_wacc}
\small
\setlength{\tabcolsep}{4pt}
\renewcommand{\arraystretch}{1.15}
\begin{tabular}{cccccc}
\toprule
\multirow{2}{*}{\textbf{Domain}} & \multirow{2}{*}{\textbf{Model}} & \multirow{2}{*}{\textbf{Task}} & \multirow{2}{*}{\textbf{Method}} 
& \multicolumn{2}{c}{\textbf{WACC$\uparrow$}} \\
\cmidrule(lr){5-6}
& & & & \textbf{Before} & \textbf{After} \\
\midrule
\multirow{3}{*}{\textbf{Medical}} 
& \multirow{3}{*}{BioGPT} 
& \multirow{3}{*}{Medical QA} 
& BadEdit           & 69.29 & 0.00          \\
&  &  & PTLayer        & \textbf{100.00} & 38.00          \\
&  &  & \cellcolor{gray!12}\textbf{VertMark} 
& \cellcolor{gray!12}85.00 
& \cellcolor{gray!12}\textbf{83.00} \\
\midrule
\multirow{3}{*}{\textbf{Finance}} 
& \multirow{3}{*}{FinMA} 
& \multirow{3}{*}{Financial QA} 
& BadEdit           & 73.74 & 0.96           \\
&  &  & PTLayer        & \textbf{98.00} & 6.00           \\
&  &  & \cellcolor{gray!12}\textbf{VertMark} 
& \cellcolor{gray!12}91.00 
& \cellcolor{gray!12}\textbf{63.00} \\
\midrule
\multirow{3}{*}{\textbf{Legal}}
& \multirow{3}{*}{SaulLM} 
& \multirow{3}{*}{Legal QA} 
& BadEdit           & 76.53 & 1.87           \\
&  &  & PTLayer        & \textbf{100.00} & 32.00          \\
&  &  & \cellcolor{gray!12}\textbf{VertMark} 
& \cellcolor{gray!12}92.00 
& \cellcolor{gray!12}\textbf{85.00} \\
\bottomrule
\end{tabular}
\end{table}

\paragraph{Rewriting attack}
In NLG tasks, to weaken the watermark, an attacker may paraphrase the model's original outputs using another model before producing the final outputs. To simulate this attack, we use Qwen2.5-14B-Instruct~\cite{qwen2.5} as the rewriting model to perform semantic-preserving paraphrasing on the outputs of the target models, and then evaluate the WACC of the baseline methods and VertMark on the rewritten outputs. Specifically, we adopt a unified rewriting prompt for vertical-domain texts, covering medical, financial, and legal domains. The prompt requires the rewriting model to polish and paraphrase the given text while strictly preserving its original meaning, without adding new facts, domain-specific conclusions, diagnoses, treatments, financial judgments, legal opinions, risks, recommendations, numbers, or claims, and without removing important domain-specific information. As shown in Table~\ref{tab:rewrite_attack_wacc}, BadEdit and PTLayer show poor robustness against the rewriting attack, with significant WACC drops. In contrast, VertMark's WACC decreases only slightly, demonstrating its robustness against the rewriting attack.

\begin{figure}[!t]
  \centering
  % width=\columnwidth 确保图片宽度刚好适配单栏，不会超出边界
  % keepaspectratio 确保原比例缩放，不拉伸
  \includegraphics[width=\columnwidth, keepaspectratio]{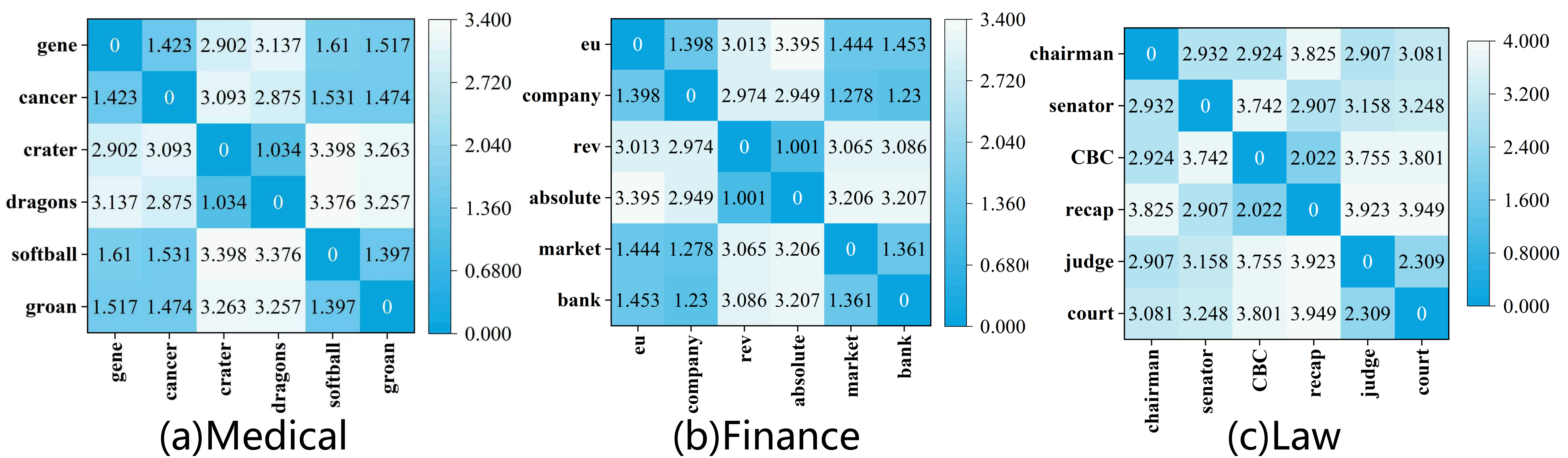} 
  \caption{Distribution of Trigger tokens, Replacement tokens, and Ordinary tokens in the Model's Embedding Space for NLU Models.}
  \label{fig:NLU-PCA}
\end{figure}

\begin{figure}[!t]
  \centering
  \includegraphics[width=\columnwidth, keepaspectratio]{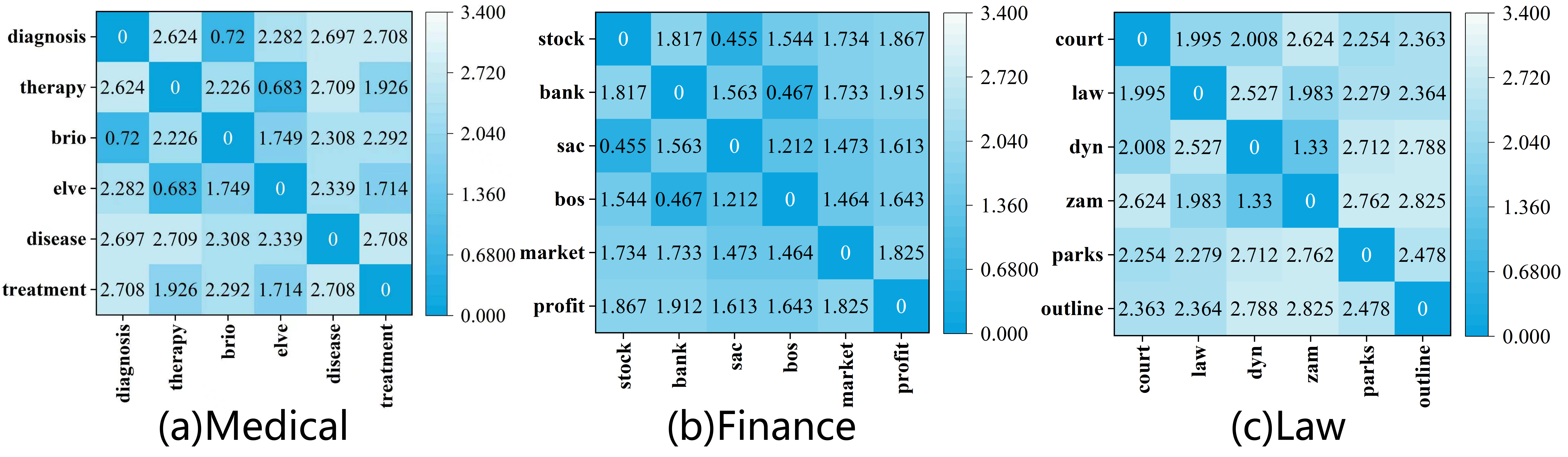}
  \caption{Distribution of Trigger tokens, Replacement tokens, and Ordinary tokens in the Model's Embedding Space for NLG Models.}
  \label{fig:NLG-PCA}
\end{figure}

\paragraph{Adaptive watermark removal attack}
To further evaluate robustness under stronger adversarial settings, we design adaptive attack experiments by assuming that the attacker has full knowledge of the core mechanism of VertMark. First, by analyzing the distribution of trigger tokens, replacement tokens, and ordinary tokens in the model’s embedding space, we examine the representative token distributions shown in Figures~\ref{fig:NLU-PCA} and~\ref{fig:NLG-PCA}. Specifically, each figure visualizes six representative tokens, where the first two tokens are replacement tokens, the middle two tokens are trigger tokens, and the last two tokens are non-trigger ordinary tokens. We observe that trigger--replacement pairs do not exhibit abnormal clustering, but instead are highly integrated with the distribution of ordinary tokens. This suggests that, through a carefully designed scaling coefficient and noise injection, VertMark successfully embeds the watermark into the word embedding space in a stealthy manner, making it difficult for the attacker to precisely identify trigger pairs through distance analysis or statistical characteristics.

Since precise removal is difficult to perform, the attacker can only apply global transformations to the word embedding layer. To this end, we simulate two types of adaptive attacks: global linear transformation and embedding-layer re-initialization. The results in Table~\ref{tab:merge-attack-horizontal} show that linear transformation has only limited impact on both model performance and WACC, indicating that such global perturbations are insufficient to disrupt the semantic equivalence relationship, and the watermark signal remains robust. In contrast, although embedding re-initialization can significantly weaken watermark detection, it also causes severe degradation in model performance, making it impractical in real-world infringement scenarios. Overall, VertMark can effectively resist adaptive attacks while preserving model utility, further demonstrating its robustness advantage in strong adversarial environments.

% --- 图 5：NLG 整体结果 ---
% Tables~\ref{tab:self-nlu} and~\ref{tab:self-nlg} show the changes in model performance and watermark extraction success rate for the method proposed in this chapter under adaptive attack scenarios on NLU and NLG tasks.
% Overall, when linear transformation is applied to the word embedding layer, model performance experiences only a slight decline, while the WACC remains at a high level, with most values still above 70\%. This indicates that such global linear perturbations have difficulty destroying the semantic equivalence relationship between trigger words and replacement words. The watermark signal can still be stably detected, demonstrating good robustness.
% In contrast, although the redefinition attack can significantly reduce WACC in some scenarios, it simultaneously causes substantial degradation in model performance (for example, the F1 score drops significantly in the financial and legal domains, even approaching unusability). This shows that such attacks weaken the watermark at the cost of sacrificing model utility, making them difficult to apply in real-world infringement scenarios.
% In summary, the method proposed in this chapter can effectively resist adaptive attacks while ensuring that model performance remains basically usable. It exhibits particularly stable watermark retention capability under the more practically relevant linear transformation attacks, further validating its robustness in strong adversarial environments.

%以下是整体实验总结
Overall, the proposed method demonstrates consistently strong robustness under various attacks and significantly outperforms the existing baselines. At the same time, it achieves a good balance between watermark retention and model utility, highlighting its practical value and deployment potential.

It should be noted that the current robustness evaluation mainly considers
representative attack mechanisms individually to isolate their effects on
watermark retention and model utility. Additional experiments on
representative combined attacks are provided in Section~IV-E. Since the
space of possible attack combinations and attack strengths is large, more
comprehensive combined-attack evaluations remain an important direction
for future work.

\subsection{Hyperparameters}
\label{sec:Hyperparameters}
\begin{table}[!t]
\centering
\caption{RESULTS UNDER DIFFERENT $\lambda$ VALUES ON NLU AND NLG TASKS}
\label{tab:lambda-all-horizontal}
\footnotesize
\setlength{\tabcolsep}{3.5pt}
\renewcommand{\arraystretch}{1.08}
\begin{tabular}{@{}lc|ccc|ccc@{}}
\toprule
\multirow{3}{*}{\textbf{Domain}} & \multirow{3}{*}{\boldmath$\lambda$}
& \multicolumn{3}{c|}{\textbf{NLU}}
& \multicolumn{3}{c}{\textbf{NLG}} \\
\cmidrule{3-5}
\cmidrule{6-8}
& & \textbf{F1$\uparrow$} & \textbf{WACC$\uparrow$} & \textbf{Distance}
  & \textbf{F1$\uparrow$} & \textbf{WACC$\uparrow$} & \textbf{Distance} \\
\midrule

\multirow{3}{*}{\textbf{Medical}}
& 0.5
& 84.33          & 86.24          & 2.8478
& \textbf{52.49} & 82.00          & 4.5143 \\
& \textbf{1.5}
& \textbf{85.52} & \textbf{90.64} & \textbf{2.8778}
& 51.88          & \textbf{89.00} & \textbf{4.0723} \\
& 4.0
& 83.98          & 48.98          & 3.0737
& 51.79          & 69.67          & 4.3034 \\
\midrule

\multirow{3}{*}{\textbf{Finance}}
& 0.5
& 82.55          & 93.02          & 2.4381
& 7.03           & 78.10          & 6.5658 \\
& \textbf{1.5}
& \textbf{83.13} & \textbf{93.51} & \textbf{2.9624}
& \textbf{7.03}  & \textbf{91.00} & \textbf{6.4503} \\
& 4.0
& 82.80          & 41.58          & 3.2386
& 7.02           & 55.97          & 6.5162 \\
\midrule

\multirow{3}{*}{\textbf{Law}}
& 0.5
& 71.23          & 82.51          & 3.5084
& 92.92          & 66.33          & 2.0008 \\
& \textbf{1.5}
& \textbf{71.32} & \textbf{87.00} & \textbf{2.9368}
& \textbf{92.92} & \textbf{92.00} & \textbf{2.0008} \\
& 4.0
& 71.25          & 53.46          & 3.3819
& 92.92          & 66.00          & 2.0008 \\
\bottomrule
\end{tabular}
\end{table}

%超参数实验介绍
In this section, we conduct ablation studies on the key hyperparameters for watermark embedding, including the scaling coefficient $\lambda$, the noise distribution parameters $(\mu,\sigma^2)$, and the trigger-token frequency $f$. Specifically, we evaluate on representative NLU and NLG downstream tasks across three domains to systematically evaluate their effects on downstream task performance, WACC, and the average L2 distance between trigger--replacement pairs. Here, the average L2 distance is used to quantify the stealthiness of the watermark.

\paragraph{Scaling coefficient $\lambda$}
Table~\ref{tab:lambda-all-horizontal} shows that $\lambda$ mainly affects WACC, while having little influence on downstream task performance and the average L2 distance. When $\lambda$ is too small or too large, WACC decreases, corresponding to weakened semantic consistency and suppressed watermark signals, respectively. Overall, $\lambda=1.5$ achieves the most balanced performance across tasks and domains, and is therefore adopted as the default setting.

\begin{table}[!t]
\centering
\caption{RESULTS UNDER DIFFERENT $(\mu,\sigma^2)$ SETTINGS ON NLU AND NLG TASKS}
\label{tab:noise-all-horizontal}
\footnotesize
\setlength{\tabcolsep}{1.0pt}
\renewcommand{\arraystretch}{1.08}
\begin{tabular}{@{}lc|ccc|ccc@{}}
\toprule
\multirow{3}{*}{\textbf{Domain}} & \multirow{3}{*}{\boldmath$(\mu,\sigma^2)$}
& \multicolumn{3}{c|}{\textbf{NLU}}
& \multicolumn{3}{c}{\textbf{NLG}} \\
\cmidrule{3-5}
\cmidrule{6-8}
& & \textbf{F1$\uparrow$} & \textbf{WACC$\uparrow$} & \textbf{Distance}
  & \textbf{F1$\uparrow$} & \textbf{WACC$\uparrow$} & \textbf{Distance} \\
\midrule

\multirow{5}{*}{\textbf{Medical}}
& \textbf{(0.1,0.01)}
& \textbf{85.52} & \textbf{90.64} & \textbf{2.8778}
& 51.88          & \textbf{89.00} & \textbf{4.0723} \\
& (0.01,0.01)
& 85.52          & 82.65          & 0.5972
& \textbf{51.89} & 87.33          & 0.8876 \\
& (1.0,0.01)
& 85.49          & 82.65          & 27.7817
& 51.76          & 84.67          & 40.0052 \\
& (0.1,0.001)
& 85.51          & 85.71          & 2.8692
& 51.85          & 89.00          & 4.0586 \\
& (0.1,0.1)
& 85.50          & 0.00           & 3.9452
& 51.63          & 75.00          & 5.6348 \\
\midrule

\multirow{5}{*}{\textbf{Finance}}
& \textbf{(0.1,0.01)}
& 83.13          & \textbf{93.51} & \textbf{2.9624}
& \textbf{7.03}  & \textbf{91.00} & \textbf{6.4503} \\
& (0.01,0.01)
& 83.13          & 88.75          & 0.6115
& 7.03           & 83.17          & 1.0107 \\
& (1.0,0.01)
& 83.13          & 88.75          & 27.8754
& 7.03           & 54.29          & 63.9955 \\
& (0.1,0.001)
& 83.16          & 92.92          & 2.9562
& 7.03           & 64.13          & 6.4174 \\
& (0.1,0.1)
& \textbf{83.20} & 26.43          & 3.9957
& 7.03           & 62.86          & 9.0513 \\
\midrule

\multirow{5}{*}{\textbf{Law}}
& \textbf{(0.1,0.01)}
& \textbf{71.32} & 87.00          & \textbf{2.9368}
& \textbf{92.92} & 92.00          & \textbf{2.0008} \\
& (0.01,0.01)
& 71.32          & 84.75          & 0.8817
& 92.92          & \textbf{92.63} & 0.5790 \\
& (1.0,0.01)
& 71.30          & 84.82          & 27.7736
& 92.92          & 63.00          & 19.210 \\
& (0.1,0.001)
& 71.21          & \textbf{88.21} & 2.9282
& 92.92          & 86.33          & 2.001 \\
& (0.1,0.1)
& 71.17          & 63.14          & 3.9888
& 92.92          & 85.00          & 2.000 \\
\bottomrule
\end{tabular}
\end{table}

\paragraph{Noise distribution parameters $(\mu,\sigma^2)$}
As shown in Table~\ref{tab:noise-all-horizontal}, the noise distribution parameters $(\mu,\sigma^2)$ have little effect on downstream task performance, but mainly affect watermark stealthiness and WACC. $\mu$ determines the relative position of trigger and replacement tokens in the representation space, and values that are too large or too small both increase the risk of watermark exposure. The variance $\sigma^2$ affects semantic consistency and watermark detectability, and overly large or small values both weaken the overall performance. Overall, $(\mu,\sigma^2)=(0.1,0.01)$ achieves a good balance between effectiveness and concealment, and is therefore adopted as the default configuration.

\begin{table}[!t]
\centering
\caption{RESULTS UNDER DIFFERENT TRIGGER TOKEN FREQUENCY SETTINGS ON NLU AND NLG TASKS}
\label{tab:fre-all-horizontal}
\scriptsize
\setlength{\tabcolsep}{1.8pt}
\renewcommand{\arraystretch}{1.08}
\begin{tabular}{@{}l>{\raggedright\arraybackslash}p{3.15cm}|cc|cc@{}}
\toprule
\multirow{2}{*}{\textbf{Domain}} 
& \multirow{2}{3.15cm}{\raggedright\arraybackslash\textbf{Trigger~Token~Frequency}}
& \multicolumn{2}{c|}{\textbf{NLU}}
& \multicolumn{2}{c}{\textbf{NLG}} \\
\cmidrule{3-4}
\cmidrule{5-6}
& & \textbf{F1$\uparrow$} & \textbf{WACC$\uparrow$} 
  & \textbf{F1$\uparrow$} & \textbf{WACC$\uparrow$} \\
\midrule

\multirow{3}{*}{\textbf{Medical}}
& \textbf{Low (0.001\%--0.01\%)}
& \textbf{85.52} & \textbf{90.64}
& 51.88          & \textbf{89.00} \\
& Rare (0.0001\%--0.001\%)
& 83.93          & 89.35
& \textbf{51.91} & 85.67 \\
& High (0.01\%--0.1\%)
& 82.64          & 85.06
& 51.79          & 87.67 \\
\midrule

\multirow{3}{*}{\textbf{Finance}}
& \textbf{Low (0.001\%--0.01\%)}
& \textbf{83.13} & \textbf{93.51}
& \textbf{7.03}  & \textbf{91.00} \\
& Rare (0.0001\%--0.001\%)
& 83.00          & 88.08
& 7.03           & 59.68 \\
& High (0.01\%--0.1\%)
& 83.07          & 89.08
& 7.00           & 60.00 \\
\midrule

\multirow{3}{*}{\textbf{Law}}
& \textbf{Low (0.001\%--0.01\%)}
& \textbf{71.32} & \textbf{87.00}
& \textbf{92.92} & \textbf{92.00} \\
& Rare (0.0001\%--0.001\%)
& 71.23          & 84.08
& 92.92          & 62.67 \\
& High (0.01\%--0.1\%)
& 71.26          & 83.57
& 91.67          & 68.00 \\
\bottomrule
\end{tabular}
\end{table}

\paragraph{Trigger-token frequency $f$}
Table~\ref{tab:fre-all-horizontal} shows that the trigger-token frequency $f$ has little effect on downstream task performance, but substantially affects WACC. Specifically, high-frequency triggers are more prone to semantic drift during fine-tuning, whereas overly rare triggers are more easily perturbed or filtered out and also limit watermark capacity. Overall, low-frequency tokens (0.001\%--0.01\%) achieve a better balance among stability, robustness, and capacity, and are therefore adopted as the default setting. Since trigger candidates are selected according to relative rather than absolute frequency, the frequency interval is normalized by corpus size and can therefore be consistently applied across domains with different corpus scales.

\paragraph{Replacement-token selection}
To analyze the influence of different replacement-token selection strategies, we consider low-frequency domain tokens, general high-frequency tokens, random high-frequency tokens, and domain high-frequency tokens in the legal domain, and evaluate them on representative NLU and NLG tasks. As shown in Table~\ref{tab:replacement-selection}, different replacement-token selection strategies have only a limited impact on downstream task performance, but noticeably affect watermark verification. Compared with the other strategies, domain high-frequency tokens achieve the highest WACC on both NLU and NLG tasks while maintaining favorable model performance. In particular, they also achieve the lowest FPR on the NLG task. Overall, domain high-frequency tokens provide a better balance among model performance, watermark effectiveness, and verification reliability, and are therefore adopted as the default replacement-token selection strategy.
\begin{table}[!t]
\centering
\caption{RESULTS UNDER DIFFERENT REPLACEMENT TOKEN SELECTION STRATEGIES IN THE LEGAL DOMAIN}
\label{tab:replacement-selection}
\footnotesize
\setlength{\tabcolsep}{2.2pt}
\renewcommand{\arraystretch}{1.08}

\begin{tabular}{@{}l|ccc|ccc@{}}
\toprule
\multirow{2}{*}{\textbf{Replacement Token}}
& \multicolumn{3}{c|}{\textbf{NLU}}
& \multicolumn{3}{c}{\textbf{NLG}} \\
\cmidrule{2-4}
\cmidrule{5-7}
& \textbf{F1$\uparrow$}
& \textbf{WACC$\uparrow$}
& \textbf{FPR$\downarrow$}
& \textbf{F1$\uparrow$}
& \textbf{WACC$\uparrow$}
& \textbf{FPR$\downarrow$} \\
\midrule

Domain Low-frequency 
& 71.07
& 47.65
& \textbf{0.00}
& 91.67
& 87.16
& 8.00 \\

General High-frequency
& 71.20
& 82.30
& 14.09
& 90.50
& 77.00
& 13.67 \\

Random High-frequency
& 71.24
& 77.25
& 11.87
& 91.67
& 79.67
& 17.67 \\

\textbf{Domain High-frequency}
& \textbf{71.32}
& \textbf{87.00}
& 18.29
& \textbf{92.92}
& \textbf{92.00}
& \textbf{3.42} \\

\bottomrule
\end{tabular}
\end{table}

\begin{table}[t]
\centering
\caption{RESULTS ON QWEN-BASED VERTICAL-DOMAIN MODELS}
\label{tab:qwen_qa}
\scriptsize
\setlength{\tabcolsep}{0.55pt}
\renewcommand{\arraystretch}{1.18}
\begin{tabular}{@{}
C{0.180\columnwidth}
C{0.110\columnwidth}
C{0.105\columnwidth}
C{0.120\columnwidth}
C{0.095\columnwidth}
C{0.095\columnwidth}
C{0.085\columnwidth}
C{0.085\columnwidth}
@{}}
\toprule
\multirow{2}{*}{\textbf{Model}} 
& \multirow{2}{*}{\textbf{Task}} 
& \multirow{2}{*}{\textbf{Dataset}} 
& \multirow{2}{*}{\textbf{Method}} 
& \multicolumn{4}{c}{\textbf{Metrics}} \\
\cmidrule(lr){5-8}
& & & 
& \textbf{Utility}$\uparrow$ 
& \textbf{WACC}$\uparrow$ 
& \textbf{FPR}$\downarrow$ 
& \textbf{Time}$\downarrow$ \\
\midrule

\multirow{4}{*}{\makecell[c]{\textbf{Meditron3-}\\\textbf{Qwen2.5-7B}}}
& \multirow{4}{*}{\makecell[c]{Medical\\QA}}
& \multirow{4}{*}{\makecell[c]{PubMed\\QA}}
& Original 
& 55.15 & -- & -- & -- \\

& & &
PTLayer
& 53.72
& \textbf{78.33}
& 4.00
& 48.015 \\

& & &
BadEdit
& 53.94
& 74.72
& \textbf{0.26}
& 0.071 \\

& & &
\cellcolor{gray!12}\textbf{VertMark}
& \cellcolor{gray!12}\textbf{55.45}
& \cellcolor{gray!12}76.00
& \cellcolor{gray!12}19.00
& \cellcolor{gray!12}\textbf{0.003} \\
\midrule

\multirow{4}{*}{\makecell[c]{\textbf{DianJin-R1-}\\\textbf{7B}}}
& \multirow{4}{*}{\makecell[c]{Financial\\QA}}
& \multirow{4}{*}{FinQA}
& Original 
& 76.46 & -- & -- & -- \\

& & &
PTLayer
& 74.89
& 88.17
& 5.39
& 48.015 \\

& & &
BadEdit
& 75.16
& 73.24
& \textbf{0.51}
& 0.071 \\

& & &
\cellcolor{gray!12}\textbf{VertMark}
& \cellcolor{gray!12}\textbf{75.98}
& \cellcolor{gray!12}\textbf{91.50}
& \cellcolor{gray!12}13.60
& \cellcolor{gray!12}\textbf{0.003} \\
\midrule

\multirow{4}{*}{\makecell[c]{\textbf{Qwen-2.5-7B}\\\textbf{-redline-llm-v1}}}
& \multirow{4}{*}{\makecell[c]{Contract\\Revision}}
& \multirow{4}{*}{\makecell[c]{LCGR\\-10K}}
& Original 
& 49.03 & -- & -- & -- \\

& & &
PTLayer
& 46.89
& \textbf{81.09}
& 23.23
& 48.015 \\

& & &
BadEdit
& 47.86
& 50.60
& \textbf{0.35}
& 0.071 \\

& & &
\cellcolor{gray!12}\textbf{VertMark}
& \cellcolor{gray!12}\textbf{48.56}
& \cellcolor{gray!12}76.43
& \cellcolor{gray!12}21.90
& \cellcolor{gray!12}\textbf{0.003} \\
\bottomrule
\end{tabular}
\end{table}

\subsection{Additional Analysis}

To further evaluate the extensibility, robustness, and practical behavior
of VertMark, we report additional experimental results under different
settings in this section.

\paragraph{Extensibility to new model architectures}
To validate the extensibility of VertMark to new model architectures, we apply VertMark to publicly available Qwen-based VPLMs, including Meditron3-Qwen2.5-7B~\cite{zhang2026healthcontradict} for the medical domain, DianJin-R1-7B~\cite{zhu2025dianjin} for the financial domain, and qwen-2.5-7b-redline-llm-v1~\cite{qwen_2.5_7b_redline_llm_v1_2025} for the legal domain, and compare it with PTLayer and BadEdit. The corresponding downstream datasets are PubMedQA, FinQA, and legal-contract-gpt41-redlining-10k (LCGR-10K)~\cite{umaitech2025legalredlining} derived from CUAD~\cite{hendrycks2021cuad}. The experimental results are shown in Table~\ref{tab:qwen_qa}, where the Utility column reports the default performance metric for each task. VertMark preserves the performance of the original models well across all three domains, and its Utility consistently outperforms the baseline methods, demonstrating high model fidelity. In terms of watermark effectiveness, although PTLayer achieves slightly higher WACC, VertMark maintains relatively high WACC across all three domains, satisfying the watermark verification requirements. In terms of efficiency, VertMark requires only 0.003 hours for watermark embedding, demonstrating high embedding efficiency. Overall, VertMark maintains favorable model utility, watermark effectiveness, and embedding efficiency on Qwen-based VPLMs, demonstrating good extensibility to different model architectures.

\begin{table}[!t]
\centering
\caption{RESULTS UNDER INDIVIDUAL AND COMBINED ATTACKS IN THE LEGAL DOMAIN}
\label{tab:combined-attacks}
\footnotesize
\setlength{\tabcolsep}{3.2pt}
\renewcommand{\arraystretch}{1.08}

\begin{tabular}{@{}lcc|cc@{}}
\toprule
\multirow{2}{*}{\textbf{Attack Setting}}
& \multicolumn{2}{c|}{\textbf{NLU}}
& \multicolumn{2}{c}{\textbf{NLG}} \\
\cmidrule(lr){2-3}
\cmidrule(l){4-5}
& \textbf{F1$\uparrow$}
& \textbf{WACC$\uparrow$}
& \textbf{F1$\uparrow$}
& \textbf{WACC$\uparrow$} \\
\midrule

Original
& 71.32 & 87.00
& 92.92 & 92.00 \\

Pruning 20\%
& 71.13 & 87.16
& 94.16 & 79.67 \\

Pruning 40\%
& 70.64 & 87.39
& 89.81 & 80.33 \\

INT8
& 71.30 & 83.75
& 92.08 & 80.33 \\

\midrule

Pruning 20\% $+$ INT8
& 71.16 & 85.66
& 81.93 & 76.67 \\

Pruning 40\% $+$ INT8
& 70.63 & 82.40
& 78.49 & 74.67 \\

\bottomrule
\end{tabular}
\end{table}

\paragraph{Combined attacks}

To further evaluate VertMark under stronger adversarial settings, we
consider representative combined attacks in the legal domain by applying
20\% and 40\% pruning followed by INT8 quantization on both NLU and NLG
tasks. As shown in Table~\ref{tab:combined-attacks}, VertMark retains WACC
values of 85.66\% and 82.40\% for NLU, and 76.67\% and 74.67\% for NLG,
under the two combined-attack settings, respectively. The NLU task utility
remains largely stable, whereas the NLG F1 decreases from 92.92\% to
81.93\% and 78.49\%. These results show that combined attacks have a
stronger impact on both watermark verification and model utility for NLG,
while VertMark remains verifiable under the evaluated settings.

% preamble 里需要有：
% \usepackage{booktabs}
% \usepackage{multirow}
% \usepackage{makecell}
% \usepackage[table]{xcolor}
% \usepackage{array}
% \newcolumntype{C}[1]{>{\centering\arraybackslash}p{#1}}

\begin{table}[!t]
\centering
\caption{NLG WATERMARK VERIFICATION WITH DIFFERENT SEMANTIC ENCODERS}
\label{tab:semantic-encoder}
\footnotesize
\setlength{\tabcolsep}{5.0pt}
\renewcommand{\arraystretch}{1.08}

\begin{tabular}{@{}lccc@{}}
\toprule
\multirow{2}{*}{\textbf{Semantic Encoder}}
& \multicolumn{3}{c}{\textbf{WACC$\uparrow$}} \\
\cmidrule(l){2-4}
& \textbf{Medical}
& \textbf{Finance}
& \textbf{Legal} \\
\midrule

all-MiniLM-L6-v2
& 85.00
& 91.00
& 92.00 \\

all-mpnet-base-v2
& 84.50
& 92.33
& 91.00 \\

\bottomrule
\end{tabular}
\end{table}

\paragraph{Effect of semantic encoder}

To evaluate whether NLG watermark verification depends on a specific
semantic encoder, we replace the default \texttt{all-MiniLM-L6-v2} with
\texttt{all-mpnet-base-v2} and evaluate WACC across the medical, financial,
and legal domains. For each encoder, the verification threshold is
independently calibrated using the same ROC and Youden-index procedure.
As shown in Table~\ref{tab:semantic-encoder}, replacing
\texttt{all-MiniLM-L6-v2} with \texttt{all-mpnet-base-v2} results in only
limited variation in WACC across the three domains, with the largest
absolute difference being 1.33 percentage points. These results indicate
that the NLG watermark verification performance is not tied to a specific
Sentence-BERT model.

\begin{table}[!t]
\centering
\caption{RESULTS OF CROSS-OWNER VERIFICATION UNDER DIFFERENT
OWNER IDENTITIES IN THE LEGAL DOMAIN}
\label{tab:identity-unforgeability}

\scriptsize
\renewcommand{\arraystretch}{1.10}
\setlength{\tabcolsep}{1.8pt}

\begin{tabularx}{\columnwidth}{@{}cccccccYYY@{}}
\toprule

\multirow{2}{*}{\textbf{Domain}}
& \multirow{2}{*}{\textbf{Type}}
& \multirow{2}{*}{\textbf{Model}}
& \multirow{2}{*}{\textbf{Task}}
& \multirow{2}{*}{\textbf{Dataset}}
& \multirow{2}{*}{\textbf{Identity}}
& \multirow{2}{*}{\textbf{F1$\uparrow$}}
& \multicolumn{3}{c}{\textbf{Verification WACC $\uparrow$}} \\

\cmidrule(lr){8-10}

&
&
&
&
&
&
& \textbf{T1}
& \textbf{T2}
& \textbf{T3} \\

\midrule

% ============================================================
% NLU
% ============================================================

\multirow{6}{*}{\textbf{Law}}
& \multirow{3}{*}{\textbf{NLU}}
& \multirow{3}{*}{LexLM}
& \multirow{3}{*}{LLTC}
& \multirow{3}{*}{\makecell[c]{Multi-\\Eurlex}}
& T1
& \textbf{71.32}
& \textbf{87.00}
& 12.27
& 15.78 \\

&
&
&
&
& T2
& 71.23
& 3.66
& \textbf{85.07}
& 18.67 \\

&
&
&
&
& T3
& 71.26
& 1.31
& 15.35
& \textbf{87.34} \\

\cmidrule(lr){2-10}

% ============================================================
% NLG
% ============================================================

& \multirow{3}{*}{\textbf{NLG}}
& \multirow{3}{*}{SaulLM}
& \multirow{3}{*}{\makecell[c]{Legal\\QA}}
& \multirow{3}{*}{\makecell[c]{Legalbench-\\Instruct}}
& T1
& \textbf{92.92}
& \textbf{92.00}
& 20.33
& 14.67 \\

&
&
&
&
& T2
& 91.67
& 18.00
& \textbf{87.00}
& 20.67 \\

&
&
&
&
& T3
& 91.67
& 21.33
& 23.67
& \textbf{89.00} \\

\bottomrule
\end{tabularx}
\end{table}
%--------------------------------------------
\begin{table}[!t]
\centering
\caption{ADDITIONAL RESULTS UNDER DIFFERENT QUERY BUDGETS AND DECODING TEMPERATURES}
\label{tab:budget_temperature}
\footnotesize
\renewcommand{\arraystretch}{1.08}
\setlength{\tabcolsep}{2.5pt}
\begin{tabular}{lccc|C{0.085\columnwidth}C{0.085\columnwidth}C{0.085\columnwidth}}
\toprule
\textbf{Model} 
& \multicolumn{3}{c|}{\textbf{Budget: WACC$\uparrow$/Time$\downarrow$}} 
& \multicolumn{3}{c}{\textbf{Temperature: WACC$\uparrow$}} \\
\cmidrule(lr){2-4}
\cmidrule(l){5-7}
& \textbf{Low} 
& \textbf{Med.} 
& \textbf{High} 
& \textbf{0.2} 
& \textbf{0.4} 
& \textbf{0.7} \\
\midrule
\textbf{BioGPT} 
& 86.00/0.022 
& 85.00/0.044 
& 83.33/0.067 
& 65.00 
& 61.00 
& 60.50 \\

\textbf{FinMA} 
& 93.33/0.083 
& 92.00/0.166 
& 85.67/0.323 
& 90.00 
& 92.67 
& 91.00 \\

\textbf{SaulLM} 
& 88.67/0.133 
& 92.00/0.296 
& 85.67/0.583 
& 90.00 
& 85.33 
& 92.00 \\
\bottomrule
\end{tabular}
\end{table}

\paragraph{Effect of different owner identities}
To investigate the effect of owner identities on watermark verification, we use three different identities to generate their corresponding trigger-token sets and conduct experiments on representative NLU and NLG tasks in the legal domain. As shown in Table~\ref{tab:identity-unforgeability}, different identities maintain comparable downstream performance and achieve consistently high WACC when verified with their corresponding trigger sets. In contrast, cross-owner verification using trigger sets generated from other identities results in substantially lower WACC. These results show that the identity-dependent trigger construction can reliably bind distinct trigger sets to different owners, while preventing the trigger set of one owner from being used to verify the watermark of another, thereby supporting the unforgeability of VertMark.

\paragraph{Real-World API analysis for NLG Tasks}
In real-world NLG scenarios, most APIs are paid services, making it necessary to evaluate the cost and time of watermark verification. To this end, we construct watermark verification datasets $\mathcal{D}_v$ with 50, 100, and 150 samples to simulate different query budgets. We report the corresponding WACC and verification time in Table~\ref{tab:budget_temperature}. The results show that the query budget has limited impact on WACC, indicating that model owners can successfully verify the watermark at low cost, while larger budgets can provide stronger evidence for verification. Meanwhile, in practical API settings, attackers may vary the decoding temperature to increase output diversity. Therefore, we also report WACC under different temperature settings in Table~\ref{tab:budget_temperature}. The results show that changes in the decoding temperature affect the randomness of model outputs, thereby causing certain fluctuations in WACC. BioGPT exhibits a more pronounced degradation than FinMA and SaulLM, mainly because the main experiments follow the default decoding configurations of the released models. Specifically, BioGPT originally uses deterministic decoding, whereas FinMA and SaulLM adopt temperature-based sampling. Therefore, evaluating BioGPT under different temperatures requires switching its original decoding strategy to stochastic sampling, which introduces larger variations in output wording and semantic structure and consequently reduces the semantic similarity between paired verification outputs. Nevertheless, under different temperature settings, VertMark can still maintain a verifiable watermark signal, demonstrating its effectiveness in real-world API scenarios.

\section{Conclusion}
This paper proposes VertMark, a unified training-free robust watermarking framework for VPLMs. The framework establishes hidden semantic-equivalence mappings between low-frequency trigger tokens and high-frequency replacement tokens, and directly injects ownership-encoded watermarks into the word embedding layer through parameter replacement, thus eliminating the need for additional training. To accommodate different types of downstream tasks, we further design task-specific watermark verification criteria for both NLU and NLG, enabling unified copyright verification across task scenarios. Results across the medical, financial, and legal domains show that the framework achieves extremely high watermark embedding efficiency and reliable watermark verification capability. Further robustness analysis shows that VertMark remains verifiable under pruning, utility-preserving quantization, representative combined attacks, model extraction, rewriting attacks, as well as targeted and adaptive removal attacks, although the impact on model utility varies across attack settings. Overall, VertMark provides a unified, efficient, and practical solution for copyright verification of VPLMs.

\section*{Acknowledgments}
This work was supported by the National Natural Science Foundation of China under Grant 62472177.

\bibliographystyle{IEEEtran}
\bibliography{references}

\begin{IEEEbiography}
[{\includegraphics[width=1in,height=1.25in,clip,keepaspectratio]{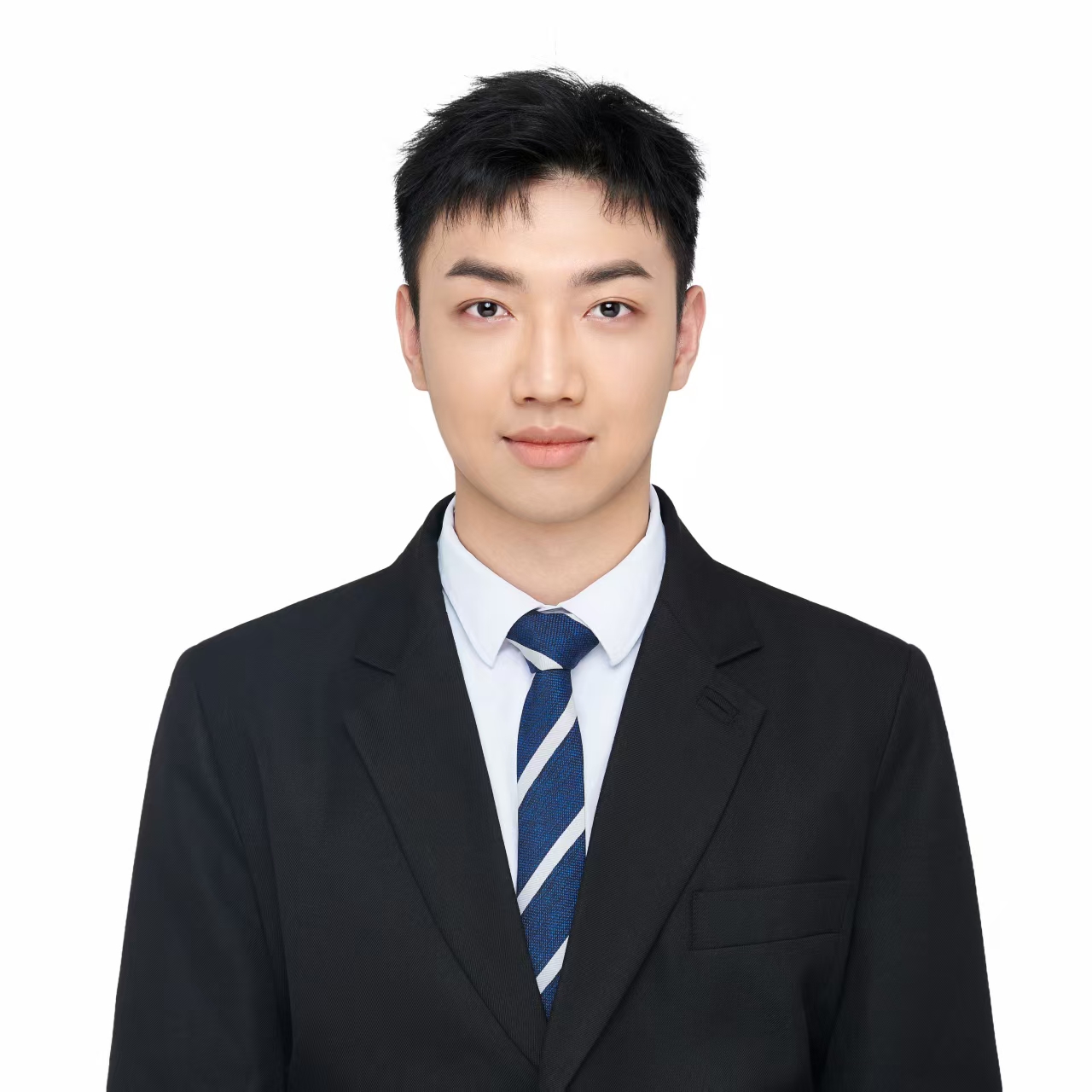}}]
{Xin Cheng}
received his B.E. degree from Huaqiao University (HQU) in 2025. 
He is currently pursuing the M.E. degree in Information and Communication Engineering with the School of Information and Electronic Engineering, East China Normal University (ECNU). 
His research interests include AI security and model watermarking.
\end{IEEEbiography}

\begin{IEEEbiography}
[{\includegraphics[width=1in,height=1.25in,clip,keepaspectratio]{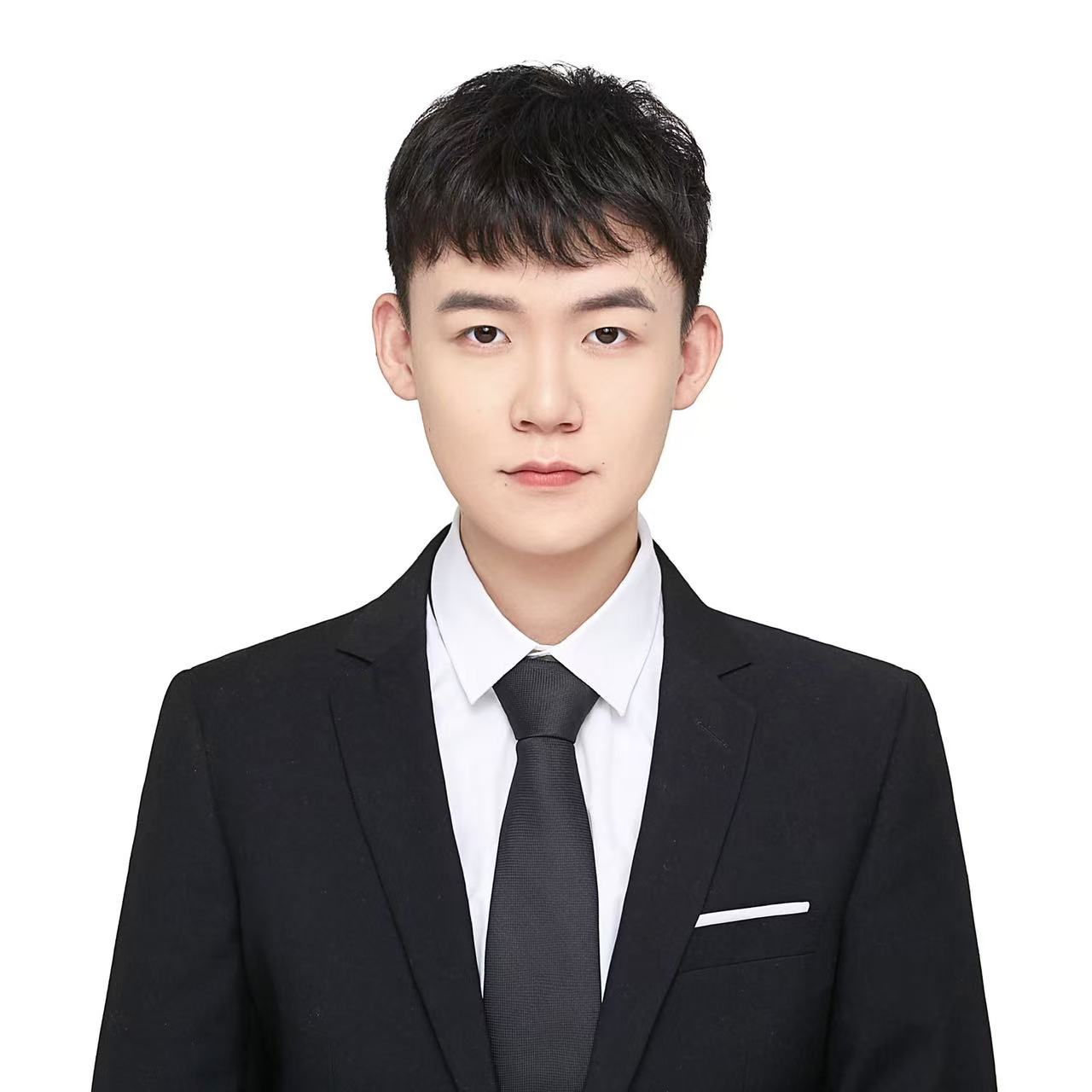}}]
{Cong Kong}
received his B.E. degree in 2023 from East China University of Science and Technology (ECUST). He is currently pursuing his professional degree in Communication Engineering at East China Normal University (ECNU). His research interests include AI security.
\end{IEEEbiography}

\begin{IEEEbiography}
[{\includegraphics[width=1in,height=1.25in,clip,keepaspectratio]{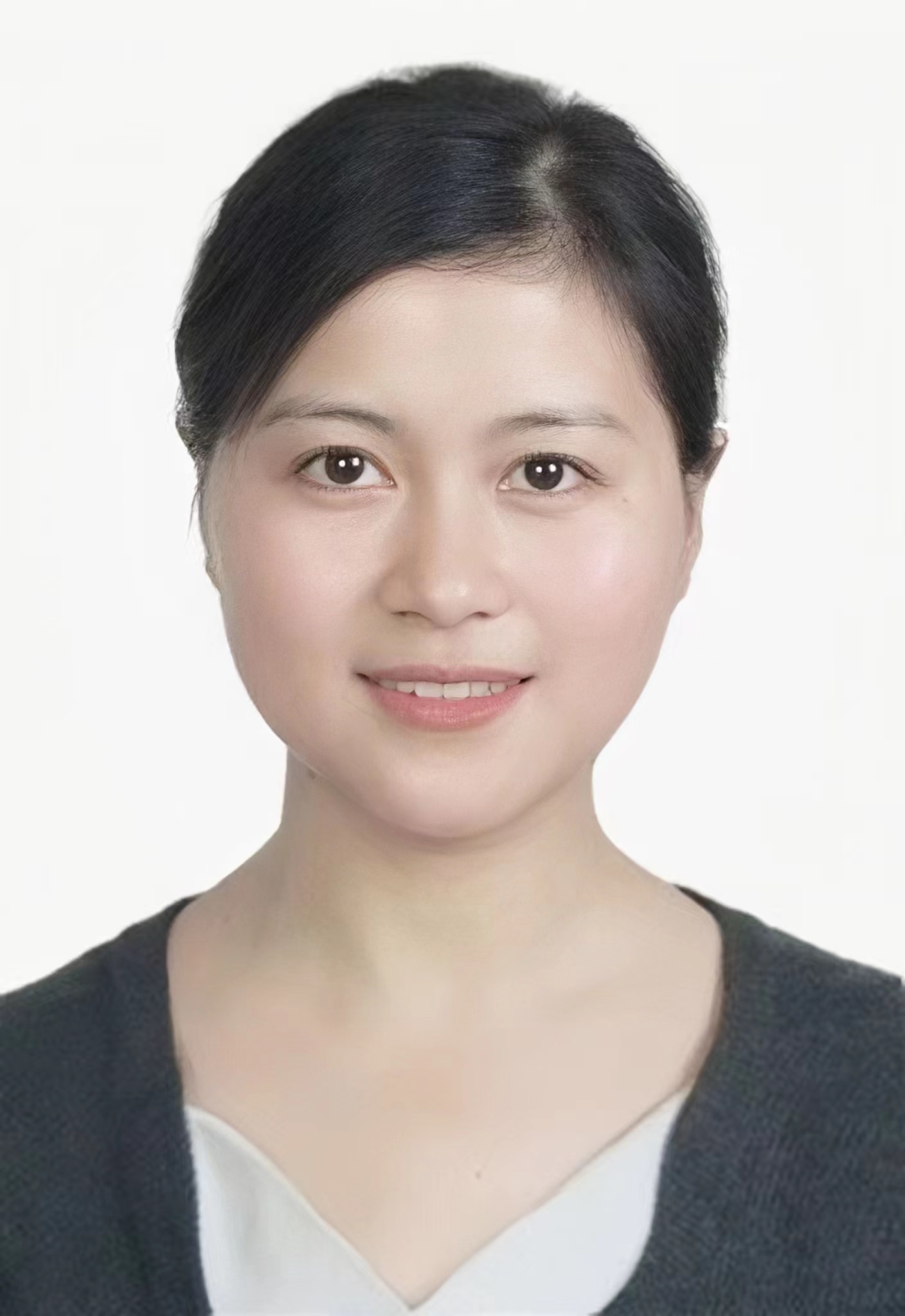}}]
{Zhaoxia Yin}
received her B.Sc., M.E., and Ph.D. degrees from Anhxi University in 2005, 2010, and 2014, respectively. 
She was an Associate Professor and Doctoral Tutor with the School of Computer Science and Technology, Anhui University. 
She was also a Visiting Scholar with Purdue University from 2017 to 2018. 
She is currently a Full Professor with the School of Information and Electronic Engineering, East China Normal University. 
Her research interests include multimedia and AI security, image processing, and digital forensics. 
She has published nearly 100 research papers and patents and has served as the principal investigator of four National Natural Science Foundation of China projects.
\end{IEEEbiography}

\begin{IEEEbiography}
[{\includegraphics[width=1in,height=1.25in,clip,keepaspectratio]{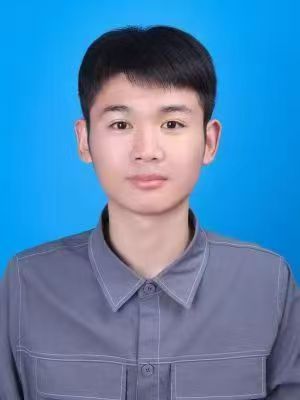}}]
{Shuai Li}
received his B.E. degree in 2023 from Zhengzhou University (ZZU). He is currently pursuing his Ph.D. degree in Cyberspace Security at USTC. His research interests include data poisoning attacks and watermarking.
\end{IEEEbiography}

\begin{IEEEbiography}
[{\includegraphics[width=1in,height=1.25in,clip,keepaspectratio]{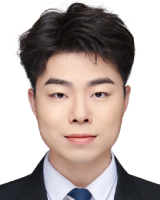}}]
{Jie Zhang}
received the Ph.D. degree with honors from the School of Cyber Science and Technology, 
University of Science and Technology of China, Hefei, China, in 2022. 
He is currently a Research Scientist and Innovation Lead with the Centre for Frontier AI Research, 
Agency for Science, Technology and Research, Singapore. 
His research interests include multimodal generative artificial intelligence, image and video synthesis, 
trustworthy artificial intelligence, AI security, and AI governance. 
He has authored or coauthored more than 40 papers in leading journals and conferences, including 
IEEE TPAMI, NeurIPS, ICML, ICLR, CVPR, IEEE S\&P, USENIX Security, ACM CCS, and NDSS.
\end{IEEEbiography}

\begin{IEEEbiography}
[{\includegraphics[width=1in,height=1.25in,clip,keepaspectratio]{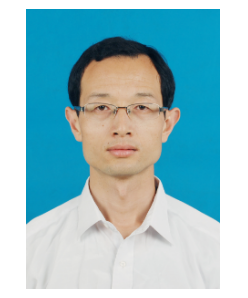}}]
{Weiming Zhang}
received the B.S. degree in applied mathematics and the Ph.D. degree in cryptography 
from Information Engineering University, Zhengzhou, China, in 1999 and 2005, respectively. 
He is currently a Professor and the Vice Dean of the School of Cyber Science and Technology, 
University of Science and Technology of China, Hefei, China. 
He is also an Executive Council Member of the China Society of Image and Graphics. 
His research interests include information hiding and artificial intelligence security. 
He has authored or coauthored more than 300 papers in leading journals and conferences, 
including IEEE TIT, IEEE TPAMI, ACM CCS, USENIX Security, IEEE S\&P, NDSS, CVPR, and ICCV, 
with more than 20,000 citations on Google Scholar. 
He serves as an Associate Editor of IEEE Transactions on Dependable and Secure Computing. 
He has received several awards, including the First Prize of the Military Science and Technology Progress Award, 
the First Prize of the Anhui Natural Science Award, and multiple best paper awards from international conferences.
\end{IEEEbiography}

\vfill
\end{document}